\documentclass[10pt,aps,floats,floatfix,amssymb,amsmath,prb,nofootinbib,twocolumn,longbibliography]{revtex4-2}

\usepackage[caption=false]{subfig}
\usepackage{physics}
\usepackage{psfrag}
\usepackage{graphicx}
\usepackage{color}
\usepackage{esint} 
\usepackage[dvipsnames]{xcolor}
\usepackage[unicode=true,pdfusetitle,
 bookmarks=true,bookmarksnumbered=false,bookmarksopen=false,
 breaklinks=false,pdfborder={0 0 0},backref=false,colorlinks=true,
 allcolors=blue]
 {hyperref}
\usepackage{geometry}
\usepackage{subfig}
\newcommand{\poc}{\hspace*{8pt}}
\newcommand{\pas}{\newline \hspace*{8pt}}

\begin{document}
\title{Dynamical quantum typicality: Simple method for investigating transport properties applied to the Holstein model}

\author{Petar Mitri\'c}
\email{mitricp@ipb.ac.rs}
\affiliation{Institute of Physics Belgrade, University of Belgrade, 
Pregrevica 118, 11080 Belgrade, Serbia}


\begin{abstract}

We investigate the transport properties of the Holstein model using the numerically exact quantum typicality (QT) approach. Roughly speaking, QT exploits the fact that even a single, randomly chosen pure state can effectively represent the full statistical ensemble in a high-dimensional Hilbert space. This allows us to compute frequency-dependent mobilities, representative of the thermodynamic limit, that are well-converged with respect to all numerical parameters. Our results are compared against other numerically exact methods,  and used to analyze the contribution of vertex corrections to frequency-dependent mobility. The promising accuracy and efficiency of the QT approach suggest its applicability to a broader class of Hamiltonians.

\end{abstract}

\maketitle

\section{\label{sec:intro} Introduction}
\poc
The study of charge carrier transport {in electron-phonon systems} is fundamental for the advancement of both the theoretical understanding and practical applications, as it directly impacts the performance of electronic devices \cite{2010_Jacoboni, 2001_Ziman}. Quantities such as the frequency dependent mobility $\mu(\omega)$  and the time-dependent diffusion constant $D(t)$ encode crucial information about transport behavior \cite{2011_Troisi, 2010_Marder}. In the weak-coupling limit, these are successfully calculated using the semiclassical Boltzmann transport equation  \cite{1965_Friedman, 1963_Glarum}. However, for stronger couplings it is necessary to go beyond the Boltzmann approach \cite{2019_Zhu, 2018_Zhou_PRL}. In principle, a pathway towards exact, and thus fully quantum, solution was laid out by Kubo in 1957 \cite{1957_Kubo} who related $\mu(\omega)$ and $D(t)$ to the current-current correlation function $C_{jj}(t)$ \cite{2000_Mahan, 2021_Bertini}. 
Yet, Kubo's approach itself does not provide a concrete recipe for calculating $C_{jj} (t)$, which is why in practice other numerically exact or approximate methods are needed for this purpose. 
\pas
Over the years, numerous approximate methods have been developed, each with a specific purpose and varying degrees of success \cite{2011_Ciuchi, 2016_Fratini, 2006_Troisi, 2024_Runeson}. Some focus exclusively on the lowest-order Feynman diagram, i.e., the bubble term~\cite{2019_Prodanovic,2022_Mitric,2003_Fratini, 2006_Fratini, 2023_Mitric, 2019_Zhu}---sometimes even with extremely high precision---but they neglect everything else, often without justification. Others approximately take into account the vertex corrections as well \cite{2008_Cheng, 2020_Fetherolf}, but face certain limitations: for example, momentum average approximation \cite{2011_Goodvin}, to the best of our knowledge, works only in the zero-temperature case; while the analytic unitary transformation \cite{2009_Ortmann} has diverging DC mobility, requiring the ad hoc introduction of some scattering time $\tau$. Even if a highly accurate approximate method emerged, one would need to demonstrate its accuracy by benchmarking it against exact results, which highlights the persistent need for numerically exact methods.
\pas
While several numerically exact methods have been developed \cite{2020_Li, 2022_Ge, 2022_Jansen, 1994_Jaklic, 2000_Jaklic, 2013_Prelovsek, 2024_Rammal, 2005_Schubert, 2006_Weiss, 2023_Miladic, 2015_Mishchenko, 2022_Wang}, they often face significant challenges, such as high computational cost due to the exponential growth of the Hilbert space with system size, or numerical instabilities like the sign problem in the quantum Monte Carlo (QMC) approach. 
This is why frequency-dependent mobility results that are both representative of the thermodynamic limit and well converged with respect to other numerical parameters are seldom encountered in the literature. 
The hierarchical equations of motion (HEOM) method \cite{2023_Jankovic, 2024_Jankovic, veljko_Pajerls_1, veljko_Pajerls_2}, capable of achieving this across a wide range of parameter regimes, has only recently been introduced. However, it also has numerical challenges, particularly in the case of strong electron-phonon couplings, where it cannot converge. In addition, it is conceptually hard (or impossible) to generalize HEOM to arbitrary electron-phonon Hamiltonian; so far, it has only been applied to Holstein \cite{2023_Jankovic, 2024_Jankovic} and Peierls \cite{veljko_Pajerls_1, veljko_Pajerls_2} models, i.e., systems with harmonic phonons and linear  electron-phonon interaction.
\pas
In this work, we apply the dynamical quantum typicality (QT) method \cite{2020_Heitmann, 2021_Jinn, 2009_Bartsch, 2018_Reimann, 2016_Steinigeweg, 2019_Richter, 2014_Monnai, 2012_Sugiura, 2013_Sugiura} to calculate transport properties in the Holstein model \cite{Holstein1959}. Within this approach, roughly speaking, the thermal expectation value from the expression for the current-current correlation function is represented using the expectation value with respect to a randomly chosen pure state, while the time evolution of $C_{jj} (t)$ is handled using the Runge-Kutta scheme \cite{2013_Elsayed,2014_Steinigeweg}. Given that QT is numerically exact in systems with a large (effective) Hilbert space dimension, which we demonstrate is typically the case in this paper, our work makes three key contributions to the existing literature:
%
%
i) We provide the frequency-dependent mobility results close to the thermodynamic limit, well converged with respect to all numerical parameters, even in the strong coupling limit. Additionally, these results are used to examine the contribution of vertex corrections to the frequency-dependent mobility (including DC mobility) in the case of strong couplings, which is the only regime not completely covered in Ref.~\cite{2024_Jankovic}.  
%
%
ii) We present the first comparison of HEOM results for optical conductivity and the current-current correlation function (for both short- and long-time dynamics) with another numerically exact method (QT), providing a more comprehensive understanding of the accuracy and performance of both methods. Previously, HEOM was only independently verified against QMC for short-time dynamics due to the QMC sign problem \cite{2023_Jankovic, 2024_Jankovic}. 
%
%
iii) Since QT is applicable to more general Hamiltonians, our study provides a crucial first step in evaluating its performance in electron-phonon systems—specifically, its numerical stability and convergence with respect to various parameters—on a well-studied system. This paves the way for applying QT to less explored systems, such as those with nonlinear electron-phonon couplings and anharmonic phonons, which have recently attracted renewed interest \cite{2023_Ragni, 2014_Berciu, 2015_Errea, 2021_Houtput, 2024_Klimin, 2023_Ranali, arxivhoutput2024_1}.
\pas
The remainder of this paper is organized as follows: Section~\ref{subsec:model} introduces the Holstein Hamiltonian. The key results from the Kubo's linear response theory are briefly summarized in Sec.~\ref{subsec:freq_dep_mob}. Section~\ref{subsec:stoc_tr_app} reviews quantum typicality and its generalization, the stochastic trace approximation, while their practical implementation for our problem is provided in Sec.~\ref{subsec_cjj}. The dynamical mean-field theory, needed for the analysis of vertex corrections, is outlined in Sec.~\ref{subsec_dmft}. In Sec.~\ref{sec:results}, we present and analyze the main QT numerical results, providing a detailed comparison with HEOM and QMC. Here, we also study the vertex corrections in the strong coupling regime. Discussion and concluding remarks are given in Sec.~\ref{Sec:discussion}. Additional numerical results are included in the Supplemental Material (SM) \cite{SuppMat}.

\vspace*{-0.2cm}
\section{\label{sec:model_method} THEORETICAL CONSIDERATIONS}
\subsection{Model} \label{subsec:model}
\vspace*{-0.3cm}
\poc
We examine the one-dimensional Holstein model, on a lattice with $N$ sites, with periodic boundary conditions. The corresponding Hamiltonian is given by
\begin{align} \label{eq:hamiltonian}
    H =& -t_0 \sum_{i} \left( c_i^\dagger c_{i+1} + c_{i+1}^\dagger c_{i} \right) \nonumber \\
      &-g\sum_i c_i^\dagger c_i \left( a_i^\dagger + a_i \right)
      +\omega_0 \sum_i a^\dagger_i a_i,
\end{align}
where $c_i^\dagger$ and $c_i$ ($a_i^\dagger$ and $a_i$) are electron (phonon) creation and annihilation operators, $t_0$ is the hopping parameter, $g$ is the coupling constant, $\omega_0$ is the dispersionless phonon frequency, while it is customary to also introduce the dimensionless strength of the electron-phonon interaction as $\lambda = g^2 / (2\omega_0 t_0)$. Throughout this paper, we set the Planck constant $\hbar$, the Boltzmann constant $k_B$, and $t_0$ to $1$. Furthermore, we assume that there is only a single electron in the conduction band. This is a standard assumption, as this model should qualitatively describe weakly doped semiconductors. 
%
\pas
%
The corresponding Hilbert space is infinite-dimensional due to the possibly infinite number of phononic excitations. Nevertheless, as shown in Sec.~A of SM \cite{SuppMat}, it is justified to truncate the Hilbert space by limiting the total number of phonons in a system to be less or equal than some (possibly large but) finite number $M$. The dimension of the corresponding Hilbert space is then given by
\begin{equation} \label{eq:dimension_of_hilbert_space}
    d = N \binom{N+M}{N}.
\end{equation}
\subsection{Frequency-dependent mobility and the current-current correlation function} \label{subsec:freq_dep_mob}
\poc
The central quantity of this work is frequency dependent mobility $\mu(\omega)$, which is just the optical conductivity normalized to the concentration of charge carriers. According to Kubo linear response theory \cite{1957_Kubo, 2000_Mahan}, $\mu(\omega)$ is related to the current-current correlation function $C_{jj}(t)$ as follows 
\begin{align} \label{eq:muw_kubo}
    \mu(\omega) &= \frac{2\tanh \left( \frac{\beta\omega}{2} \right)}{\omega}
    \int_0^\infty \cos(\omega t) \, \mathrm{Re}\;C_{jj}(t) \, \mathrm{d}t, \\
    C_{jj}(t) &= \langle j(t) j \rangle = 
    \frac{\mathrm{Tr}\left[ e^{-\beta H} e^{iHt} j e^{-iHt} j \right]}{\mathrm{Tr}\left[ e^{-\beta H} \right]}.
    \label{eq:curr_curr_corr_f}
\end{align}
Here, \(\beta = 1/T\) is the inverse temperature, while \(\langle \dots \rangle = \mathrm{Tr}[e^{-\beta H} \dots] / \mathrm{Tr}[e^{-\beta H}]\) represents the thermal expectation value. The symbol \(j\) represents the current operator, which for the Holstein model reads as  \cite{2023_Jankovic, 2000_Mahan}
\begin{equation} \label{eq:current_op}
j = i t_0 \sum_r \left( 
c^\dagger_{r+1} c_r - c_r^\dagger c_{r+1}
\right),
\end{equation}
while \(j(t) = e^{iHt} j e^{-i H t}\) represents the current operator in the Heisenberg picture. 
\pas
While $\mathrm{Re}\,C_{jj}(t)$ contains all information about $\mu(\omega)$, it is sometimes \cite{2016_Fratini, 2024_Jankovic} more convenient to analyze an equivalent quantity: the time dependent diffusion constant $D(t)$, which describes the growth rate  ${D(t) = \frac{1}{2} \frac{d}{dt} \big(\Delta x(t)\big)^2}$ of the charge carrier quantum-mechanical spread, $\Delta x(t) = \sqrt{\langle (x(t) - x(0))^2 \rangle}$. The diffusion constant is mathematically related to  $\mathrm{Re}\,C_{jj}(t)$ via the following expression
\begin{equation} \label{eq:diffusion_const}
    D(t) = \int_0^t dt' \, \mathrm{Re} \,C_{jj}(t').
\end{equation}
From Eqs.~\eqref{eq:muw_kubo}~and~\eqref{eq:diffusion_const}, it directly follows that ${\mu(\omega=0) = D(t\to\infty) /T}$. This is the famous Einstein relation, which shows that for the DC mobility $\mu(\omega=0)$ to be finite, the time-dependent diffusion constant must saturate to a constant value at large times.
\pas
Before explaining how we calculate $C_{jj}(t)$ in practice, let us first emphasize that $\mu(\omega)$ satisfies the so-called optical sum rule, which in the case of the Holstein model in the thermodynamic limit $(N\to\infty)$ reads as follows
\begin{equation} \label{eq:op_sum_rule}
    \int_{0}^\infty  \mathrm{d}\omega \, \mu(\omega) = -\frac{\pi}{2} \langle H_\mathrm{el} \rangle
    =-\frac{\pi}{2} \frac{\mathrm{Tr}\left[ e^{-\beta H} H_\mathrm{el} \right]}{\mathrm{Tr}\left[ e^{-\beta H} \right]}
    .
\end{equation}
Here, $\langle H_\mathrm{el} \rangle$ is the thermal expectation value of the electronic part of the Hamiltonian, which is given by the first line of Eq.~\eqref{eq:hamiltonian}. On a finite lattice $N$, Eq.~\eqref{eq:op_sum_rule} is violated, which is why the relative error
%
\begin{equation} \label{eq:delta_op_sum_rl}
    \delta = \frac{|\int_{0}^\infty  \mathrm{d}\omega \, \mu(\omega) +\frac{\pi}{2} \langle H_\mathrm{el} \rangle|}{-\frac{\pi}{2} \langle H_\mathrm{el} \rangle},
\end{equation}
 indicates the presence of finite-size effects \cite{2023_Jankovic, 2000_Jaklic}.
%
\pas

\vspace*{-0.8cm}
\subsection{Calculating traces using stochastic trace estimation and quantum typicality methods}\label{subsec:stoc_tr_app}
\vspace*{-0.2cm}
%
%
\poc
As the Hilbert space in our system is typically large, we need an efficient way to calculate the traces in Eqs.~\eqref{eq:curr_curr_corr_f}~and~\eqref{eq:op_sum_rule}. This can be achieved using the so-called stochastic trace estimation \cite{2017_Saibaba, 1989_Girard, 1989_Hutchinson, 2000_Hams, 2004_Iitaka, hutch++, xtrace}.
In short, using mutually independent complex random variables \( c_n \), with zero mean and arbitrary variance \( E\left[ |c|^2 \right] \), the trace of arbitrary operator \( X \) can be obtained as follows \cite{2021_Jinn}
\begin{align} \label{eq:aux:r43r}
    \mathrm{Tr} X &= 
    \sum_{n,m} \delta_{n,m} \langle n | X | m \rangle 
    = \sum_{n,m} \frac{E\left[ c_n^* c_m \right]}{E \left[ |c|^2 \right]} \langle n | X | m \rangle
    \nonumber \\
    &= \frac{1}{E \left[ |c|^2 \right]} E \left[
    \left( \sum_n c_n^* \langle n| \right) X \left( \sum_m c_m |m \rangle \right)
    \right] \nonumber \\
    &= \frac{E \left[ \langle \psi | X | \psi \rangle \right]}{E\left[ |c|^2 \right]}.
\end{align}
Here, $|n \rangle$ are (arbitrary) basis vectors, the symbol $E[\dots]$ denotes an expectation value with respect to the random numbers $c_n$, while $|\psi\rangle  = \sum_n c_n |n\rangle$ can be interpreted as a randomly chosen vector from the Hilbert space. 
While various probability distributions for the random coefficients $c_n$ are commonly used in the literature \cite{2021_Jinn, 2006_Weiss}, this work will use the Gaussian distribution.
\pas
Setting $X=e^{-\beta H}A$ in Eq.~\eqref{eq:aux:r43r}, and using the cyclic property of the trace  $\mathrm{Tr}\left[e^{-\beta H} A \right] = \mathrm{Tr}\left[e^{-\beta H/2} A e^{-\beta H/2} \right]$, we see that the expectation value of an arbitrary quantity $A$ in the canonical ensemble can be expressed as
\begin{equation} \label{eq:stoch_tr_appr}
     \langle A \rangle = 
     \frac{E \left[ \langle \psi | e^{-\beta H/2} A e^{-\beta H/2} | \psi \rangle \right]}{E \left[ \langle \psi | e^{-\beta H} | \psi \rangle \right]} = 
     \frac{E\left[ \langle \psi_\beta | A| \psi_\beta \rangle \right]}{E\left[ \langle \psi_\beta | \psi_\beta \rangle \right]},
\end{equation}
where $| \psi_\beta \rangle = e^{-\beta H /2} | \psi \rangle$. 
%
If expectation values $E[.]$ in Eq.~\eqref{eq:stoch_tr_appr} are approximated by using only $R$ random vectors $|\psi\rangle$, the relative error scales as $\mathcal{O}\left(1/\sqrt{R\, d_\mathrm{eff}} \right)$, where $d_\mathrm{eff}=\mathrm{Tr}\left[ e^{-\beta (H - E_0)} \right]$ is the effective dimension of the Hilbert space, and $E_0$ is the ground state energy; see Ref. \cite{2020_Schnack} and Sec.~B from the SM \cite{SuppMat}. Consequently, in cases where the effective Hilbert space is large, as it typically is in this work, a single random vector suffices for  Eq.~\eqref{eq:stoch_tr_appr}. This approach is known as (dynamical) quantum typicality (QT) \cite{2020_Heitmann, 2021_Jinn}.
%
%
%

\vspace*{-0.4cm}
\subsection{Quantum typicality approach for computing current-current correlation functions} \label{subsec_cjj}
\vspace*{-0.4cm}
%
\poc
Let us now demonstrate how QT, which we summarized in the previous subsection, can be applied for the calculation of the current-current correlation function $C_{jj}(t) = \langle j(t) j \rangle$. Starting from Eq.~\eqref{eq:stoch_tr_appr}, with only a single random vector $|\psi\rangle$, and setting $A=j(t) j= e^{i H t} j e^{- i H t} j$, we see that  $C_{jj}(t)$ can be expressed as
\begin{align} \label{eq:cjj_corr_f}
    C_{jj}(t) &\approx
     \frac{\langle\psi_\beta (t) | j | \phi_\beta (t) \rangle}{\langle \psi_\beta(t) | \psi_\beta(t)\rangle},
\end{align}
where we introduced
\begin{align}
    |\psi_\beta(t) \rangle &= e^{-iHt} e^{-\beta H /2} | \psi \rangle, \label{eq:psi_beta} \\
    |\phi_\beta(t) \rangle &= e^{-iHt} j e^{-\beta H /2} | \psi \rangle. \label{eq:phi_beta}
\end{align}

To deal with $e^{-i H t}$ ($e^{-\beta H}$ is handled analogously), we  break the total evolution into small time steps $dt$
\begin{equation}
    |\psi(t) \rangle = e^{-iHt} | \psi \rangle = 
    \underbrace{e^{-iH dt}e^{-iH dt}\dots e^{-iH dt}}_{N_t \, \text{times, such that }N_t dt = t}
    |\psi\rangle,
\end{equation}
and then expand each $e^{-i H dt}$ in the Taylor series up to some order $n_{RK}$ (in this paper we usually choose $n_{RK}=4$)
\begin{equation} \label{eq:small_time_evolution}
    | \psi(dt) \rangle \equiv e^{-i H dt} | \psi \rangle \approx \sum_{l=0}^{n_{RK}} 
    \frac{(-iH dt)^l}{l!} | \psi \rangle.
\end{equation}
In practice, Eq.~\eqref{eq:small_time_evolution} is implemented by introducing
\begin{equation} \label{eq:reccurence_relations_fj}
    | f_s \rangle = \sum_{l=0}^{n_{RK}-s} \frac{s!}{(s+l)!} \left( -i H dt \right)^l | \psi \rangle,
\end{equation}
which satisfy the following recurrence relations
\begin{equation} \label{eq:reccurence_relations_fj_22345}
    |f_{s-1} \rangle = | \psi \rangle  + \frac{(-i H dt)}{s} | f_s \rangle,
\end{equation}
with a property that $| f_{n_{RK}} \rangle =  | \psi \rangle$ and $ | f_0 \rangle = | \psi(dt) \rangle$. Therefore, the expression in Eq.\eqref{eq:small_time_evolution} can be evaluated as follows: one starts with $| f_{n_{RK}} \rangle = | \psi \rangle$ and applies Eq.~\eqref{eq:reccurence_relations_fj_22345} to iteratively calculate $|f_s\rangle$ for $s = {n_{RK}-1}, {n_{RK}-2}, \dots, 0$, until $| \psi(dt) \rangle = | f_0 \rangle$ is obtained. This approach is commonly referred to as the Runge-Kutta method \cite{2013_Elsayed, 2014_Steinigeweg, 2020_Heitmann}. It has the property that for the time propagation of $|\psi\rangle$, it requires storing only two $d$-dimensional vectors—one for $|\psi(t) \rangle$ and the other as an auxiliary vector. Although Eqs.~\eqref{eq:cjj_corr_f}--\eqref{eq:phi_beta} involve the propagation of two vectors ($|\psi\rangle$ and $|\phi\rangle$), the auxiliary vector can be shared between them, so only three $d$-dimensional vectors are needed in total. It is worth noting that the Runge-Kutta scheme does not necessarily preserve the normalization of the wave function $|\psi(dt)\rangle$, which is why we manually adjust $|\psi(dt)\rangle$ at each time step by multiplying it with a real constant, ensuring that $\langle \psi| \psi \rangle = \langle \psi(t)| \psi(t)\rangle$.
%
%
\pas
This approach, combining stochastic trace estimation (with a single random vector) and the Runge-Kutta scheme,
will be referred to as the quantum typicality (QT) for brevity. QT can also analogously be used to compute static quantities such as $\langle H_\mathrm{el}  \rangle$. In fact, $C_{jj}(t)$ and $\langle H_\mathrm{el} \rangle$  can both be calculated simultaneously using QT, by sharing the same random vector $|\psi\rangle$.
%
%


%
\vspace*{-0.5cm}
\subsection{Dynamical mean field theory} \label{subsec_dmft}
\vspace*{-0.3cm}
\poc
QT can provide numerically exact results for \( \mu(\omega) \), but to assess the importance of vertex corrections, we also need results from the bubble approximation. These can be directly obtained using the single-particle Green's (or spectral) functions. In the case of the Holstein model, the dynamical mean field theory (DMFT) provides a highly accurate way to calculate these quantities \cite{2022_Mitric}.
\pas
DMFT is a nonperturbative method \cite{1996_Georges} designed to tackle problems with local interaction, such as the Hubbard or Holstein models \cite{1997_Ciuchi}, by relating them to the impurity problem with a self-consistency relation. In general, it is exact in the case when the crystal lattice is infinite dimensional, while it is otherwise approximate, as it assumes a momentum-independent self-energy. However, the special case of the Holstein model brings two important simplifications: i) DMFT is extremely accurate for arbitrary dimensionality of lattice, even for 1D \cite{2022_Mitric}. ii) the impurity problem has an analytic solution, in the form of continuous fraction expansion \cite{1997_Ciuchi}. In addition, DMFT can be applied directly in the thermodynamic limit $N\to\infty$. As a consequence, DMFT is here perfectly suitable for the numerically efficient and accurate calculation of  \( \mu(\omega) \) within the bubble approximation. For more details, see Refs. \cite{2022_Mitric, 2024_Jankovic}.
%


\vspace*{-0.3cm}
\section{\label{sec:results}Results and analysis}
\vspace*{-0.1cm}
\poc
Here, we calculate the transport properties within QT, across various parameter regimes, and compare the results with the predictions from other numerically exact methods: HEOM and QMC. We note that the HEOM data were taken from Refs. \cite{2024_Jankovic, 2023_Jankovic,Zenodo_Jankovic}, and the QMC data from Ref. \cite{2024_Jankovic}, while the methodological frameworks underlying these approaches are detailed in Refs. \cite{2023_Jankovic} and \cite{2023_Miladic}, respectively. In addition, we also perform the DMFT calculations, but only for a strong electron-phonon coupling, as this is the only regime where vertex corrections were not previously analyzed in Ref. \cite{2024_Jankovic}. Before presenting the main results in Sec.~\ref{sec:main_results}, we first discuss potential sources of inaccuracies within QT in Sec.~\ref{sec:conv_analysis_12}, which can arise in practice despite QT being in principle exact.


\vspace*{-0.6cm}
\subsection{Convergence analysis of QT with respect to various numerical parameters}
\vspace*{-0.3cm}
\label{sec:conv_analysis_12}
\poc
To obtain accurate QT results representative of the thermodynamic limit, the following conditions must be satisfied:
 (i) results must fully converge with  respect to the number of phonons $M$ in the Hilbert space; we have demonstrated how this is checked in practice in Sec.~A of SM \cite{SuppMat}. (ii) Results must fully converge with respect to the number of lattice sites $N$.
 While this is often established by checking that the optical sum rule is satisfied to a high degree of accuracy \cite{2005_Schubert} (the quantity $\delta$, given by Eq.~\eqref{eq:delta_op_sum_rl}, is displayed for each of the Figs.~\ref{Fig:weak_intermediate_w0_1}--~\ref{Fig:adiabatic_strong}), such test is strictly speaking only a necessary condition for the result to be representative of the thermodynamic limit.  Instead, it is much better to simply calculate the transport properties for various $N$, and directly inspect whether the results actually converged; this was done Sec.~C of SM \cite{SuppMat}.
 %
 (iii) A sufficiently small time step $dt$ must be used in the Runge-Kutta method; for all the regimes we examined we checked that $dt=0.01$ is sufficiently small. An example of such analysis can be seen in Fig.~S7 of SM \cite{SuppMat} (iv) The fact that reliable results can be obtained using only a single random vector in stochastic trace approximation; this was checked both in Sec.~B of SM \cite{SuppMat}, and in Sec.~\ref{sec:main_results} of the main text as well. Namely, in all Figs.~\ref{Fig:weak_intermediate_w0_1}--~\ref{Fig:adiabatic_strong} (usually in insets) we show two QT results, for two different lattice sites $N_\mathrm{larger}^\mathrm{QT}$ and $N_\mathrm{smaller}^\mathrm{QT}$. Since these two results use different random vectors $|\psi\rangle$, their agreement would demonstrate both that finite-size effects are small and that using a single random vector in the stochastic trace approximation is sufficient.
\pas
Although all of the above conditions were satisfied for the regimes presented in Figs.~\ref{Fig:weak_intermediate_w0_1}–\ref{Fig:adiabatic_strong}, we note that there are some regimes where satisfying these conditions could require impractically large computational resources. For example, in the weak coupling regime, electron-phonon scattering is also weak, meaning the electron must travel long distances to lose memory of its initial state; this is even more pronounced at low temperatures. Consequently, large number of lattice sites $N$ are required to obtain results representative of the thermodynamic limit. For such large $N$, possibly involving several dozen lattice sites, if we consider that there is only a single phonon per lattice site $M=N$, the dimension of the Hilbert space (as seen from Eq.~\eqref{eq:dimension_of_hilbert_space}) is enormous as it scales as $d \sim 4^N \sqrt{N/\pi}$. This explains why QT, in its present form, is not suitable for weak couplings and low temperatures, and why we focus on intermediate and strong couplings at moderate to high temperatures.

\subsection{Numerical results} \label{sec:main_results}
\subsubsection{Weak-intermediate and intermediate couplings}
\begin{figure}[!t]
  \includegraphics[width=3.4in,trim=0cm 0cm 0cm 0cm]{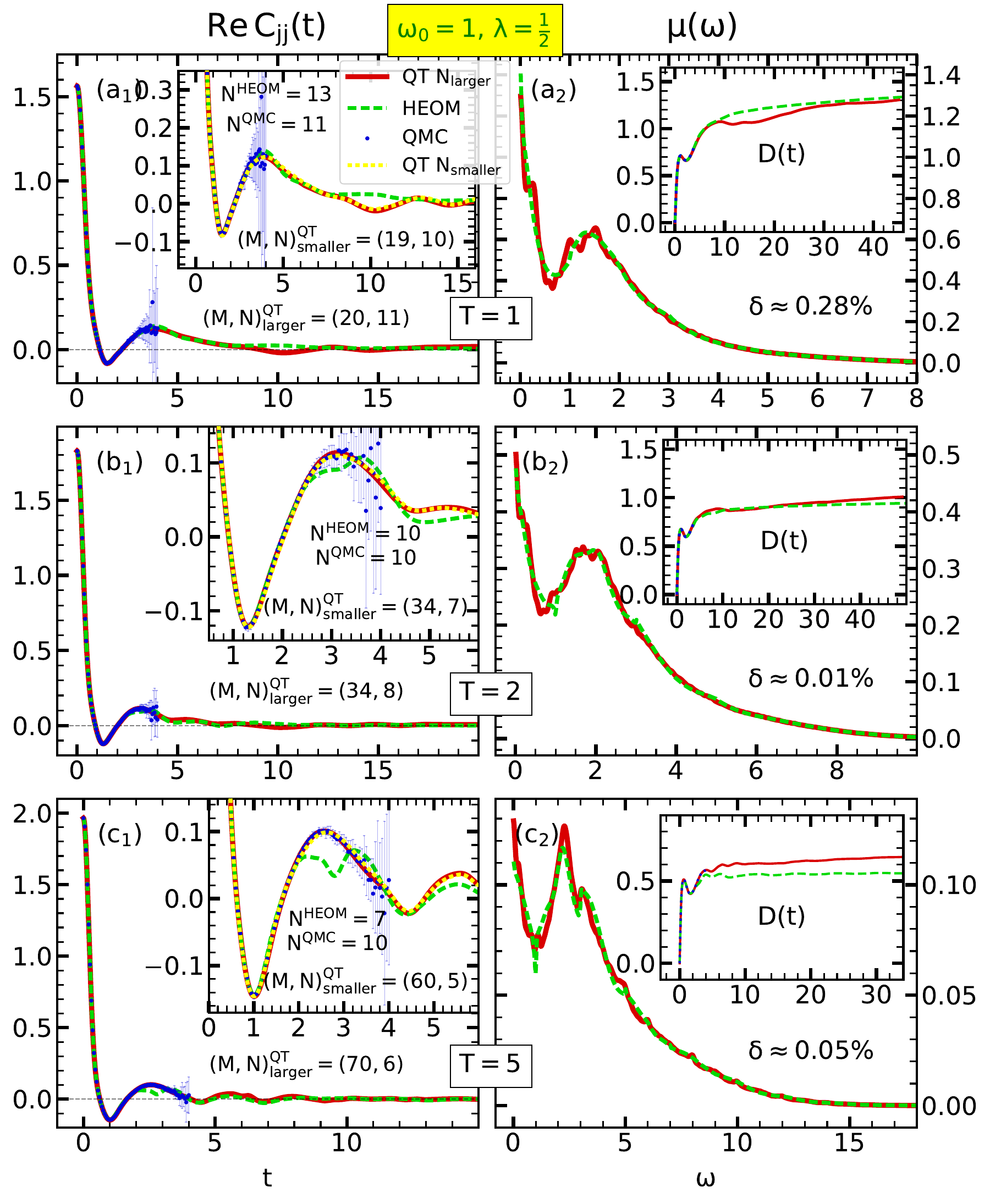} 
 \caption{$\!\!$Comparison of $(\mathrm{a}_1)$--$(\mathrm{c}1)$ $\mathrm{Re} \, C_{jj}(t)$ and $(\mathrm{a}_2)$--$(\mathrm{c}_2)$ $\mu(\omega)$ in the weak-intermediate coupling regime ${\omega_0=1}$, $\lambda=1/2$ at $T=1,2,5$. Insets of $(\mathrm{a}_1)$--$(\mathrm{c}_1)$ show zoomed-in portions of the panels and additional QT results on smaller lattices. Panels $(\mathrm{a}_2)$--$(\mathrm{c}_2)$ display the relative accuracy $\delta$ of the optical sum rule within QT, with insets showing $D(t)$.  
 In all panels, QMC results are presented with the corresponding error bars.
 }
 \label{Fig:weak_intermediate_w0_1}
\end{figure}
\begin{figure}[!t]
  \includegraphics[width=3.4in,trim=0cm 0cm 0cm 0cm]{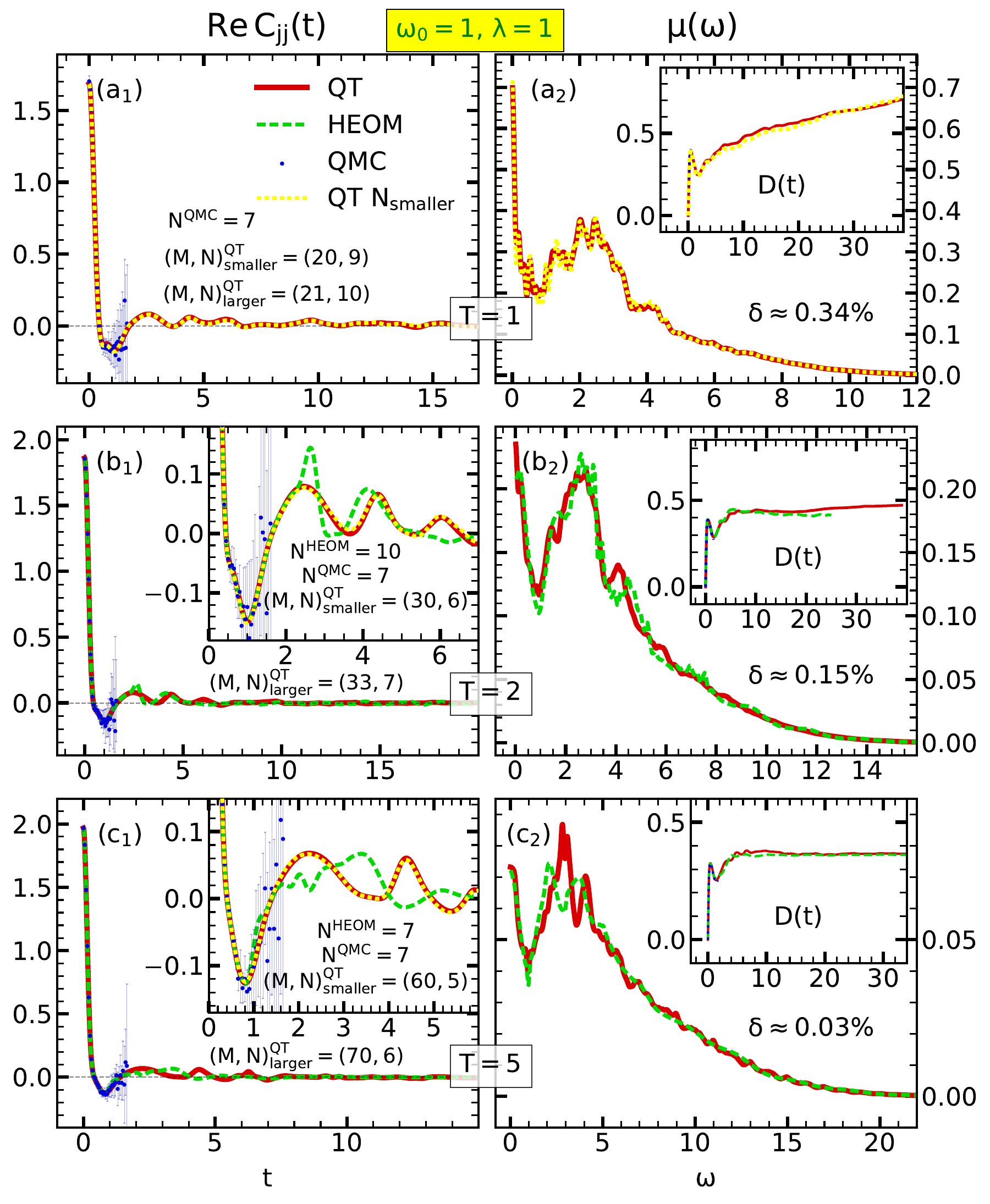} 
 \caption{
  Comparison of $(\mathrm{a}_1)$--$(\mathrm{c}1)$ $\mathrm{Re} \, C_{jj}(t)$ and $(\mathrm{a}_2)$--$(\mathrm{c}_2)$ $\mu(\omega)$ in the intermediate coupling regime $\omega_0=1$, $\lambda=1$ at $T=1,2,5$. Insets of $(\mathrm{a}_1)$--$(\mathrm{c}_1)$ show zoomed-in portions of the panels and additional QT results on smaller lattices. Panels $(\mathrm{a}_2)$--$(\mathrm{c}_2)$ display the relative accuracy $\delta$ of the optical sum rule within QT, with insets showing $D(t)$. 
  QMC results for $N^\mathrm{QMC} = 7$ are presented in all panels, along with their corresponding error bars.
 }
 \label{Fig:intermediate_w0_1}
\end{figure}
\begin{figure}[!t]
  \includegraphics[width=3.4in,trim=0cm 0cm 0cm 0cm]{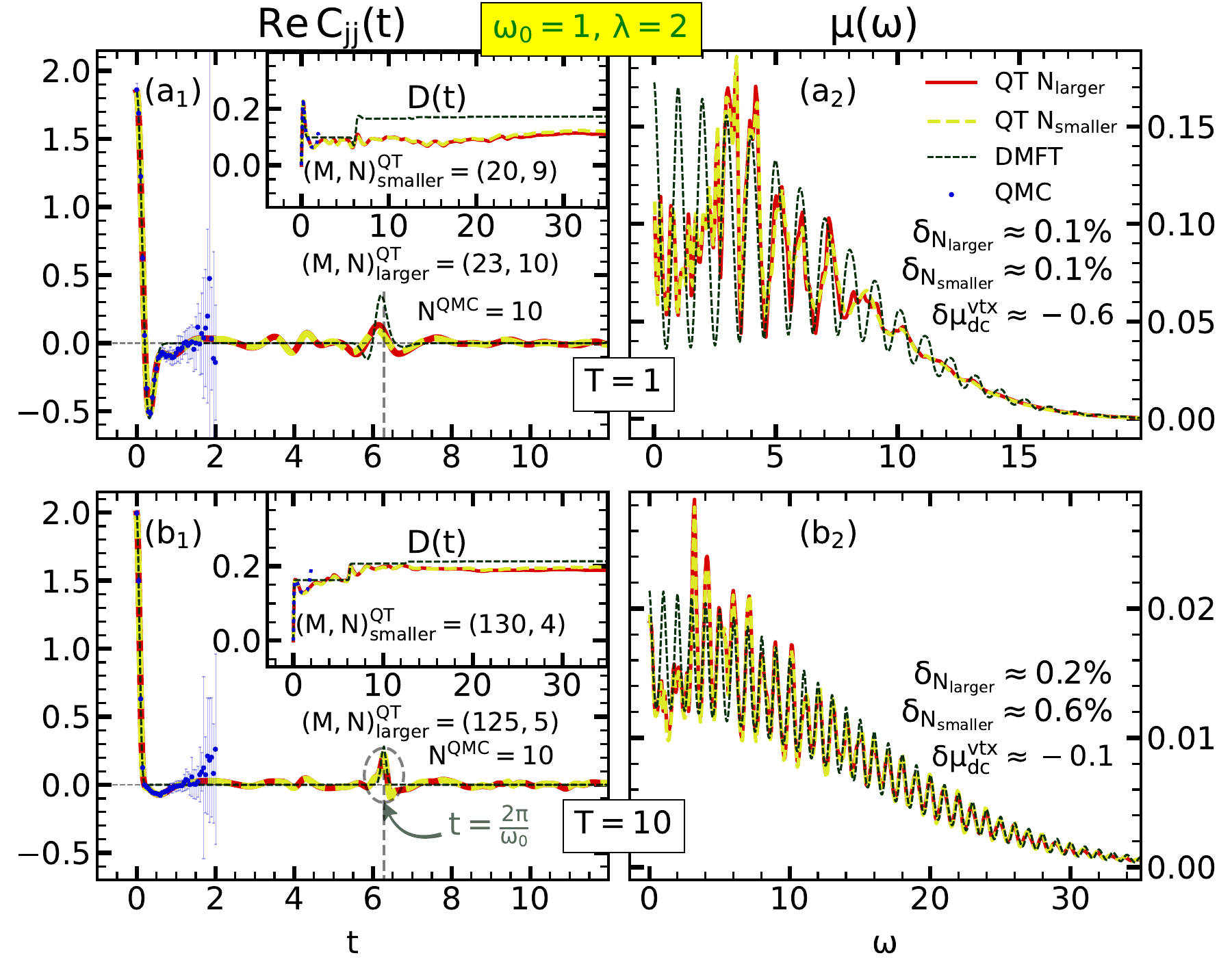} 
 \caption{Comparison of $(\mathrm{a}_1)$--$(\mathrm{b}_1)$ $C_{jj}(t)$ and $(\mathrm{a}_2)$--$(\mathrm{b}_2)$ $\mu(\omega)$ in the strong coupling regime $\omega_0 = 1$, $\lambda = 2$, at $T=1, 10$. 
 Insets of $(\mathrm{a}_1)$--$(\mathrm{b}_1)$ show $D(t)$, while $\delta$ (within QT) and $\delta\mu_\mathrm{dc}^\mathrm{vtx}$, given by Eqs.~\eqref{eq:delta_op_sum_rl}~and~\eqref{eq:significance_vtx_corr}, are displayed in Panels $(\mathrm{a}_2)$--$(\mathrm{b}_2)$.  
 QMC results are presented in all panels with the corresponding error bars.
 }
 \label{Fig:strong_w0_1}
 \vspace*{-0.7cm}
\end{figure}
\begin{figure}[!t]
  \includegraphics[width=3.4in,trim=0cm 0cm 0cm 0cm]{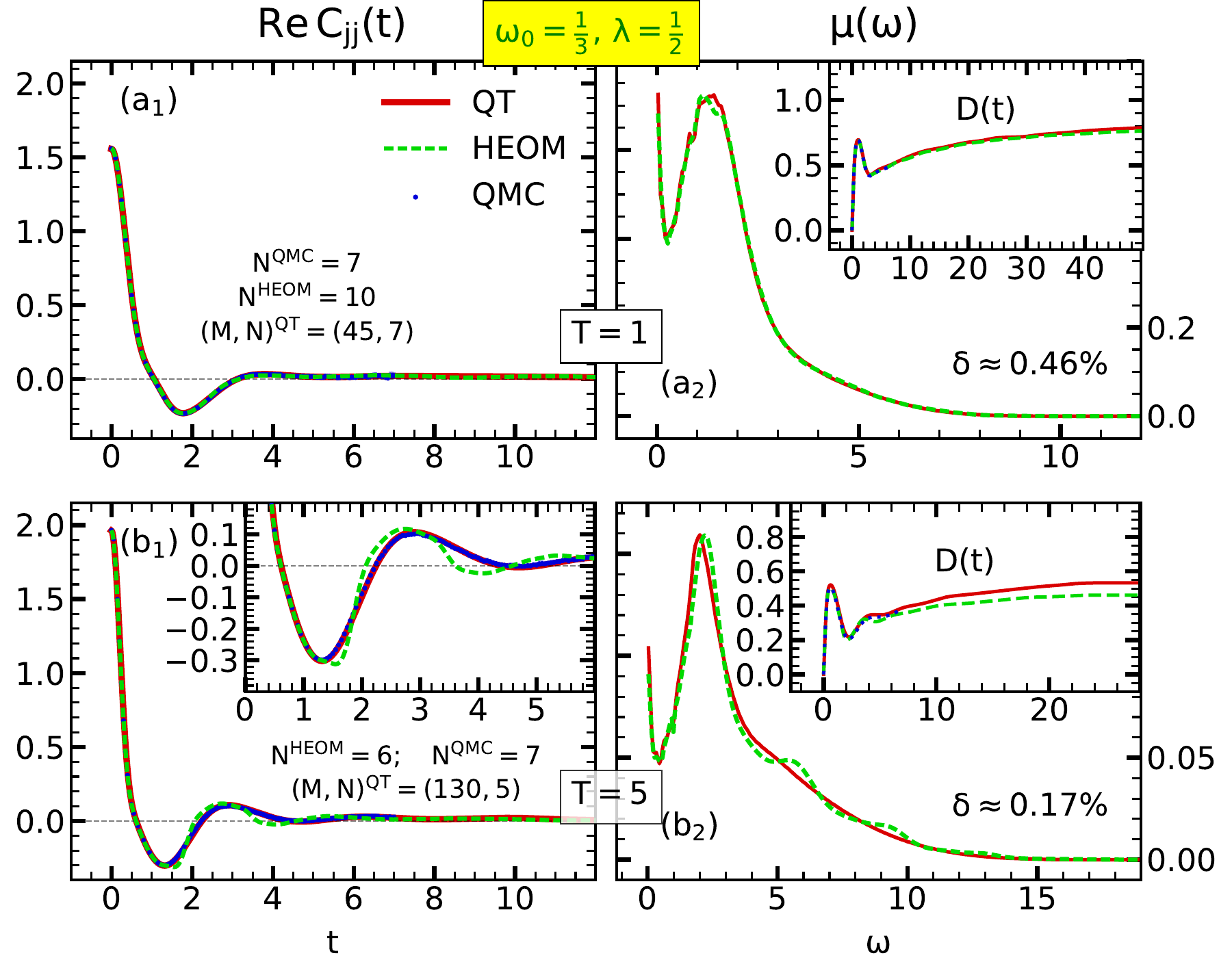} 
 \caption{Comparison of $(\mathrm{a}_1)$--$(\mathrm{b}_1)$ $C_{jj}(t)$ and $(\mathrm{a}_2)$--$(\mathrm{b}_2)$ $\mu(\omega)$ for $\omega_0 = 1/3$, $\lambda = 1/2$, at $T=1, 5$. Inset of $(\mathrm{b}_1)$ displays the zoomed-in portion of the corresponding Panel. In $(\mathrm{a}_2)$--$(\mathrm{b}_2)$ we also show $\delta$ (within QT), given by Eq.~\eqref{eq:delta_op_sum_rl}, as well as the time-dependent diffusion constant $D(t)$ in the insets.  
 Error bars for the QMC results  are smaller than the symbol size.
 }
 \label{Fig:adiabatic_weak}
\end{figure}
\begin{figure}[!t]
  \includegraphics[width=3.4in,trim=0cm 0cm 0cm 0cm]{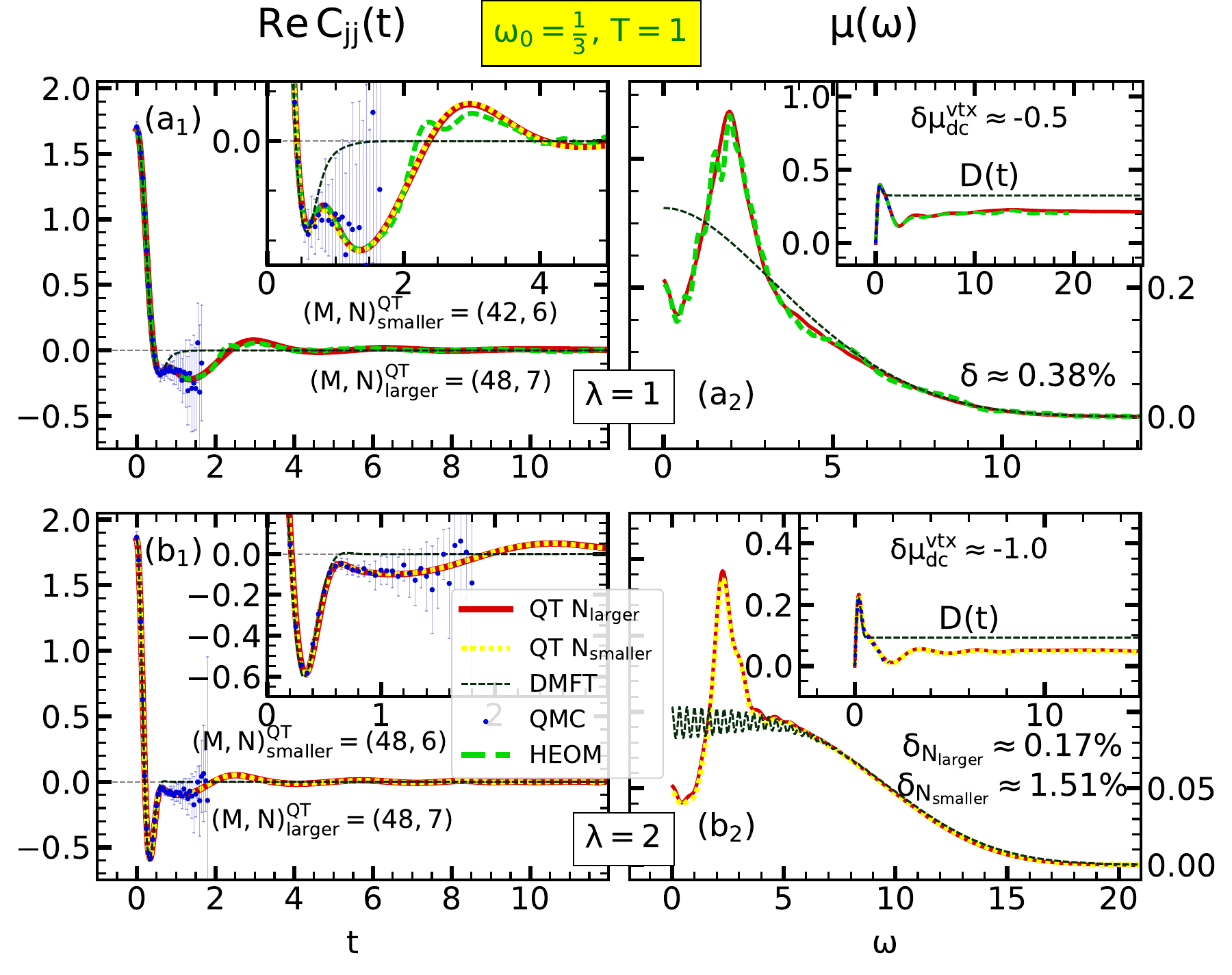} 
 \caption{Comparison of $(\mathrm{a}_1)$--$(\mathrm{b}_1)$  $C_{jj}(t)$ and $(\mathrm{a}_2)$--$(\mathrm{b}_2)$ $\mu(\omega)$ for $\omega_0=1/3$, $T=1$ and $\lambda=1,2 $.  Inset of $(\mathrm{a}_1)$--$(\mathrm{b}_1)$ display zoomed-in portions of corresponding Panels. In $(\mathrm{a}_2)$--$(\mathrm{b}_2)$ we also show $\delta$ (within QT), given by Eq.~\eqref{eq:delta_op_sum_rl}, as well as the time-dependent diffusion constant $D(t)$ in the insets. 
  QMC results for $N^\mathrm{QMC} = 7$ are presented in all panels, along with their corresponding error bars.
 }
 \label{Fig:adiabatic_strong}
\end{figure}

The results for weak-intermediate coupling regime $\lambda=1/2$, moderate temperature $T=1$ and phonon frequency $\omega_0=1$ are presented in Figs.~\ref{Fig:weak_intermediate_w0_1}$(\mathrm{a}_1)$~and~\ref{Fig:weak_intermediate_w0_1}$(\mathrm{a}_2)$. In Fig.~\ref{Fig:weak_intermediate_w0_1}$(\mathrm{a}_1)$ we observe an excellent agreement between all methods. A small discrepancy in the current-current correlation function $C_{jj}(t)$ between QT and HEOM  is visible only for $t\sim 10$. To analyze this discrepancy, let us note once again that QT results here (and also for all regimes presented in Figs.~\ref{Fig:weak_intermediate_w0_1}–\ref{Fig:adiabatic_strong}) satisfy all convergence conditions from Sec.~\ref{sec:conv_analysis_12}: we made sure that the Runge-Kutta time step is sufficiently small; we checked in Fig.~S1 of the SM \cite{SuppMat} that $\mathrm{Re}\;C_{jj}(t)$ has fully converged with respect to the number of phononic excitations $M$;  the agreement of QT results for $N_\mathrm{larger}$ and $N_\mathrm{smaller}$ in the inset of Fig.~\ref{Fig:weak_intermediate_w0_1}$(\mathrm{a}_1)$ demonstrate both that the results are well converged with respect to the number of lattice sites $N$ (see also Fig.~S5 from the SM \cite{SuppMat}), as well as the fact that sufficiently accurate results can be obtained using only a single random vector in the stochastic trace approximation. 
This analysis, together with the simplicity and transparency of the QT method, give us confidence that QT is indeed more accurate in this case. In contrast to QT, the HEOM method, although very powerful, involves more complex numerical settings and is less straightforward, making it harder to determine whether small numerical inaccuracies have been fully eliminated \cite{2024_Jankovic, 2023_Jankovic}. Despite all of this, we find that the agreement in frequency dependent mobility $\mu(\omega)$ is excellent; see Fig~\ref{Fig:weak_intermediate_w0_1}$(\mathrm{a}_2)$. 
%
%
%
\pas
As the temperature increases, the discrepancies between QT and HEOM grow progressively more pronounced, becoming evident even at intermediate timescales; see Figs.~\ref{Fig:weak_intermediate_w0_1}$(\mathrm{b}_1)$--\ref{Fig:weak_intermediate_w0_1}$(\mathrm{c}_1)$. At $T=5$, a clear difference in $C_{jj}(t)$, between QT and HEOM can be seen in an inset, showing a closer view of Panel~\ref{Fig:weak_intermediate_w0_1}$(\mathrm{c}1)$. We again assert that QT is more accurate, as all convergence conditions from Sec.~\ref{sec:conv_analysis_12} are met, a point further supported by QMC; see Fig.~\ref{Fig:weak_intermediate_w0_1}$(\mathrm{c}1)$. The reason we place such confidence in the QMC results, allowing them to arbitrate between QT and HEOM, is the presence of corresponding error bars, which explicitly show the precision of the QMC results. Present analysis, in combination with the discussion associated with Fig.~5 of Ref. \cite{2023_Jankovic}, indicates that HEOM results are probably not fully converged with respect to the so-called maximum hierarchy depth \cite{2023_Jankovic, 2024_Jankovic}. Nevertheless, the  mismatch between QT and HEOM is not very large.
%
%
%
%
%
\pas
The general agreement between the QT and HEOM current-current correlation functions, translates to a very good agreement between the frequency-dependent mobilities computed by these approaches; see Figs.~\ref{Fig:weak_intermediate_w0_1}$(\mathrm{a}_2)$--\ref{Fig:weak_intermediate_w0_1}$(\mathrm{c}_2)$. Although this confirms that QT can actually provide accurate $\mu(\omega)$, it should be noted that the quantitative value of the DC mobility (and $\mu(\omega)$ for very small $\omega$) has somewhat larger error. This is evident from the analysis of $D(t)$, keeping in mind the Einstein relation ${\mu(\omega=0) = D(t\to\infty) /T}$. As we see, a distinct plateau (i.e., the saturation) of $D(t)$ does not always emerge at the longest times over the interval for which we propagate $D(t)$, especially at lower temperatures. Longer time propagation, although numerically expensive, is possible, but we note that the finite-size effects are more pronounced for this feature. As discussed in Sec.~C of SM \cite{SuppMat}, it seems that $D(t)$ saturates more quickly for larger lattices. Therefore, HEOM results have an advantage for calculating DC mobilities, as they are obtained on larger lattices and propagated to longer times than QT.  However, the already mentioned small inaccuracies in the HEOM solution at intermediate timescales can also affect the DC mobility result. For example, by comparing the insets in Figs.\ref{Fig:weak_intermediate_w0_1}$(\mathrm{c}_1)$--\ref{Fig:weak_intermediate_w0_1}$(\mathrm{c}_2)$, we see that slight differences in the DC mobilities predicted by HEOM and QT are largely due to discrepancies in $\mathrm{Re}\;C_{jj}(t)$ at $t \sim 2.5$ and $t \sim 5.5$. Despite these differences, the DC mobilities predicted by QT and HEOM remain in good agreement across all the regimes we examined. 
%
%
%
%
%
\pas
The results for $\lambda = 1$ are shown in Fig.~\ref{Fig:intermediate_w0_1}. For $T=2,5$, we see that similar conclusions can be drawn as in $\lambda=1/2$ case, reinforcing our confidence in QT being more reliable than HEOM at higher temperatures. However, a peculiarity in the QT solution can be observed at $T=1$: DC mobility cannot be estimated reliably as the diffusion constant $D(t)$ does not exhibit any signs of saturation at long times; see the inset of Fig.~\ref{Fig:intermediate_w0_1}$(\mathrm{a_2})$. As discussed in Sec.~C of SM \cite{SuppMat}, we expect this to be a consequence of the finite-size effects. Nevertheless, $\mu(\omega)$ for $|\omega| \gtrsim 0.15$ should be quite accurate. Unfortunately, HEOM predictions for this particular regime ($T=1$) are not available.
\pas
\subsubsection{Strong couplings}
%
For $\lambda=2$ and $\omega_0=1$, the renormalized electron mass (at $T=0$) is about $10$ times larger than band mass;  see Fig.~(1b) from Ref. \cite{2022_Mitric} or Fig.~3 from Ref. \cite{1998_kornilovitch}. This is why this is considered a strong coupling regime. The corresponding transport properties, for two different temperatures, $T=1, 10$, are examined in Fig.~\ref{Fig:strong_w0_1}. 
\pas
Such strong interactions, while requiring a large number of phonons $M$, do not necessarily demand an extensive number of lattice sites $N$ to approximate the thermodynamic limit effectively. This is demonstrated in Fig.~\ref{Fig:strong_w0_1} by showing both that the optical sum rule is satisfied to a high degree of accuracy, and also by explicitly comparing QT results for two consecutive lattice sizes. This makes it quite suitable for application of QT method, having in mind how the Hilbert space dimension scales with $N$ and $M$ (see Eq.~\eqref{eq:dimension_of_hilbert_space}).
%
In contrast, HEOM method cannot be applied here due to the fact that it cannot converge with respect to the maximum hierarchy depth. Thus, the only  independent benchmark that we have is the QMC,  and we see that it is in excellent agreement with QT. Unfortunately, QMC is reliable only for relatively short times, until the sign problems starts showing up.  Nevertheless, QT should be accurate even for longer times, as it satisfies all conditions from Sec.~\ref{sec:conv_analysis_12}. A further support for this statement can be seen from certain qualitative features that the QT solution satisfies. 
%
These include rather fast attenuation of $C_{jj}(t)$, as well as the appearance of a bump at $t=2\pi / \omega_0$  which is characteristic feature of such regimes \cite{2024_Jankovic, 2023_Miladic}. 

\newpage \clearpage
\poc
In Fig.~\ref{Fig:strong_w0_1} we also show the DMFT predictions which, as we already explained in Sec.~\ref{subsec_dmft} and Ref. \cite{2024_Jankovic}, are practically the exact results without the vertex corrections. Thus, by comparing them to the QT results we can inspect exactly how large the vertex corrections actually are. 
As observed, the short-time results for \( C_{jj}(t) \) from both methods match perfectly, leading to excellent agreement in the \( \mu(\omega) \) predictions at higher frequencies, \( \omega \gtrsim 3 \). These predictions display a series of peaks at integer multiples of \( \omega_0 \), with the difference being that DMFT yields simple peaks, whereas QT produces peaks with more intricate internal structure. 
%
As the temperature is increased, we see that the discrepancy between QT and DMFT gets smaller, which is in accordance with the analytic result \cite{2024_Jankovic} that the vertex corrections are vanishing in the high-temperature limit. Let us also note that a decent saturation of the QT (and DMFT) diffusion constant is observed, meaning that DC mobility can be  estimated. This is especially true at higher temperatures $T>1$; see Sec.~D of SM \cite{SuppMat}. Thus, we are also able to measure the importance of vertex corrections for DC mobility, that we quantify by introducing
\begin{equation} \label{eq:significance_vtx_corr}
    \delta \mu_{dc}^\mathrm{vtx} = \frac{\mu_{dc}^\mathrm{QT} - \mu_{dc}^\mathrm{DMFT}}{\mu_{dc}^\mathrm{QT}},
\end{equation}
which is shown in Figs.~\ref{Fig:strong_w0_1}~and~S11 from the SM \cite{SuppMat} for $T=1,2,5,10$. All of these results readily demonstrate that the order of magnitude of the DC mobility is correctly predicted by the bubble approximation, even for strong coupling regime.

%
\subsubsection{Approaching adiabatic limit}
Near the adiabatic limit $(\omega_0 = 1/3)$, the results for the weak-intermediate $(\lambda=1/2)$, intermediate $(\lambda=1)$, and strong coupling regime $(\lambda = 2)$ are presented in Figs.~\ref{Fig:adiabatic_weak}~and~\ref{Fig:adiabatic_strong}. 
The results are analogous to those we already obtained for $\omega_0=1$: we see that the current-current correlation functions (from all methods) are in excellent agreement for $\lambda=1/2$, $T=1$, while at $T=5$ slight discrepancies between QT and HEOM solutions arise already at intermediate timescales. These are a consequence of small inaccuracies of the HEOM solution, as confirmed by the QMC results which are in perfect agreement with QT. An analogous conclusion can be also drawn for intermediate couplings $\lambda=1$ as well. For this case, however, the accuracy of the QT results was verified by carefully checking all the convergence conditions we listed in Sec.~\ref{sec:conv_analysis_12}. Despite the small differences in $C_{jj}(t)$, we find that QT and HEOM predictions for $\mu(\omega)$ agree quite well  (see Fig.~\ref{Fig:adiabatic_weak}$(\mathrm{b}_2)$~and~\ref{Fig:adiabatic_strong}$(\mathrm{a}_2)$). The results in the strong coupling regime $\lambda=2$ are available only within QT and QMC, and the agreement is excellent, while HEOM results are unavailable due to the fact that it fails to converge with respect to the maximum hierarchy depth. Let us also note that both QT results in Fig.~\ref{Fig:adiabatic_strong} display a quite good long-time saturation of the diffusion constant $D(t)$, enabling us to reliably extract the value of DC mobility. 
\pas
In Fig.~\ref{Fig:adiabatic_strong} we also show the DMFT results that, by comparing them with QT predictions, are used to examine the contribution of vertex corrections. As observed, the results differ notably from the $\omega_0 = 1$ case. Here, the bubble approximation shows a much larger qualitative deviation from the exact solution, failing to capture its most prominent feature: the displaced Drude peak around $\omega \approx 2 t_0$. This peak is a consequence of temporal localization of the electron at small timescales, that occurs due to the phonons, which act as a source of dynamical disorder in the system \cite{2016_Fratini, 2024_QC_Mitric}. Within this picture, it is expected that as we increase the interaction strength, the electron should be more localized (at short time scales), implying that the displaced Drude peak should move to larger frequencies. This is exactly what we see in Fig.~\ref{Fig:adiabatic_strong}: the displaced peak for $\lambda=1$ is centered around $\omega \approx 1.95$, while for $\lambda = 2.0$ it shifts to $\omega \approx 2.25$. Regarding the vertex corrections to DC mobility, let us note that although they are not extremely large in Fig.~\ref{Fig:adiabatic_strong} ($\delta\mu_\mathrm{dc}^\mathrm{vtx}\approx -0.5$ for $\lambda=1.0$, while $\delta\mu_\mathrm{dc}^\mathrm{vtx}\approx -1.0$ for $\lambda=2.0$), they will become much larger as we approach the adiabatic limit $\omega_0\to 0$. This is a consequence of the fact that $\mu(\omega)$ in the exact solution will tend to zero due to Anderson localization, while the DMFT result will stay finite, as it neglects the nonlocal correlations \cite{2024_QC_Mitric}.

\section{Discussion and Conclusions} \label{Sec:discussion}
\poc
In summary, we employed the stochastic trace estimation with a single random vector, combined with the Runge-Kutta method—collectively referred to as QT—to compute the transport properties of the Holstein model. This approach allowed us to converge the results with respect to all numerical parameters, yielding highly accurate (i.e., numerically exact) results representative of the thermodynamic limit. To accomplish this, it was important not to waste RAM memory by storing too many unnecessary auxiliary states; our implementation required only three $d$-dimensional vectors, allowing us to work in Hilbert spaces with large number of dimensions $d$. For such large $d$, the computational cost per random vector (from the stochastic trace approximation) is not cheap in terms of CPU time. This was why, for the calculation to be feasible, it is important 
that even a small number of (or in our case even a single) random vector was sufficient to obtain accurate results. 
\pas
We note that computational efficiency could potentially be improved by using the finite-temperature Lanczos method or the kernel polynomial method (KPM), both of which also rely on stochastic trace estimation. However, these approaches would require storing hundreds of additional auxiliary states in RAM \cite{2024_Rammal, 2005_Schubert}, which we avoided. While KPM can also be applied in a much more memory-efficient way, this comes at the cost of a significantly increased computational effort, which now scales quadratically with the expansion order \cite{2006_Weiss}. This is why we decided to simply use QT.
\pas
The obtained QT results were compared with the predictions from the HEOM method, which is also numerically exact in principle, yielding a deeper insight into the accuracy and effectiveness of both approaches. 
We found that QT can handle much stronger electron-phonon couplings, which are inaccessible to HEOM due to convergence issues with respect to the maximum hierarchy depth. Moreover, in the intermediate coupling regime at elevated temperatures—where both methods are applicable—QT yielded more accurate results for the current-current correlation function at intermediate timescales. This conclusion was drawn by carefully establishing that all potential sources of error, within the numerically exact QT framework, were eliminated.
%
\pas
Further unequivocal support for the previous statement was provided by the numerically exact real-time QMC data, which aligned perfectly with the QT results, while the corresponding QMC error bars were significantly smaller than the discrepancy between QT and HEOM. Although the QMC error was (in some regimes) small enough to arbitrate between QT and HEOM at intermediate timescales, we note that the QMC error becomes much larger at longer times due to the infamous sign problem, implying that QMC often cannot be used for the calculation of frequency-dependent mobility $\mu(\omega)$. Such numerical instabilities are not present in QT, which can be propagated to sufficiently long times for the calculation of $\mu(\omega)$. Therefore, QT effectively addresses many of the limitations of QMC and HEOM. However, it should be noted that QT does not replicate all of their advantages: in particular, it cannot reach systems as large as QMC and HEOM, which is crucial in weak coupling and low temperature regimes, where a large number of lattice sites is necessary to obtain results representative of the thermodynamic limit.
\pas
QT results were also used to examine the significance of vertex corrections to the frequency-dependent mobility $\mu(\omega)$ for strong couplings, as this is the only regime where such an analysis, due to the absence of strong coupling results, had not already been performed in Ref.~\cite{2024_Jankovic}. For $\omega_0=1$, we found that the bubble approximation results were qualitatively quite similar to the exact results. While a difference could be observed quantitatively, especially for smaller $\omega$, the bubble approximation managed to predict the correct order of magnitude, even for the DC mobility. With increasing temperature, we verified that the importance of vertex corrections reduced, in accordance with the analytic results of Ref.~\cite{2024_Jankovic}. The vertex corrections were also examined for $\omega_0=1/3$: here, even qualitatively, the bubble approximation could not capture the correct behavior of $\mu(\omega)$, missing a key feature - the displaced Drude peak, associated with a temporal localization of electrons at short timescales. 
\pas
Although our treatment in this work focused on the Holstein model, the same procedure can be readily applied to more general systems. The only adjustment required would be modifying how the Hamiltonian $H$ and the current operator $j$ act on an arbitrary vector from the corresponding Hilbert space. However, our algorithm involves numerous matrix-vector multiplications, $H|\psi\rangle$ or $j|\psi\rangle$, making it crucial to execute the procedure efficiently to maintain computational feasibility. This imposes a practical limitation on a system which we can examine, which are thus the ones where $H$ and $j$ can be represented as sparse matrices.  
This remark, along with our promising results in the case of the Holstein model, make the QT method particularly well-suited for studying systems such as the Peierls model \cite{2006_Troisi, 2009_Fratini} or systems with nonlinear electron-phonon couplings and anharmonic phonons \cite{2023_Ragni, 2014_Berciu, 2015_Errea, 2021_Houtput, 2024_Klimin, 2023_Ranali, arxivhoutput2024_1}.

\section*{Acknowledgments}
The author acknowledges useful discussions with Darko Tanaskovi\' c and Veljko Jankovi\' c, and also thanks Veljko Jankovi\' c for sharing his HEOM data from Refs. \cite{2023_Jankovic, 2024_Jankovic, Zenodo_Jankovic}, as well as Nenad Vukmirovi\' c for sharing his QMC data from Ref. \cite{2024_Jankovic}. The author
acknowledges funding provided by the Institute of Physics
Belgrade through a grant from the Ministry of Science,
Technological Development, and Innovation of the Republic of Serbia. A part of the numerical computations was performed on
the PARADOX-IV supercomputing facility at the Scientific
Computing Laboratory, National Center of Excellence for the
Study of Complex Systems, Institute of Physics Belgrade.


\section*{DATA AVAILABILITY}

The data that support the findings of this article are openly
available \cite{Zenodo_Mitric_QT}.



%


\clearpage
\pagebreak
\newpage

%


\onecolumngrid
\begin{center}
  \textbf{\large Supplemental material for: Dynamical quantum typicality: Simple method for investigating transport properties applied to the Holstein model}\\[.2cm]
  P. Mitri\'c\\[.1cm]
  {\itshape Institute of Physics Belgrade, University of Belgrade, Pregrevica 118, 11080 Belgrade, Serbia} \\[.1cm]
\end{center}

\setcounter{equation}{0}
\setcounter{figure}{0}
\setcounter{table}{0}
\makeatletter
\renewcommand{\theequation}{S\arabic{equation}}
\renewcommand{\thefigure}{S\arabic{figure}}
\renewcommand{\bibnumfmt}[1]{[S#1]}
\renewcommand{\citenumfont}[1]{S#1}
\renewcommand{\thetable}{S\arabic{table}}

In the main part of the paper, we used the (dynamical) quantum typicality (QT) method \cite{SM_2020_Heitmann, SM_2021_Jinn} to calculate the transport properties in the 1D Holstein model and we compared the obtained results with the predictions of the hierarchical equations of motion (HEOM) method, quantum Monte Carlo (QMC), and dynamical mean-field theory (DMFT). In particular, we presented the current-current correlation functions $C_{jj}(t)$, alongside related quantities such as optical conductivities $\mu(\omega)$ and time-dependent diffusion constants $D(t)$, in a wide range of parameter regimes. In this Supplemental Material, we present some additional numerical results and discussions, complementing those in the main text.

\section{Convergence with respect to the total number of phonons $M$ in the Hilbert space}
\poc
Since the QT methodology is limited to finite-dimensional Hilbert spaces, we confined our study to systems with $N$ lattice sites and restricted the total number of phonons to some finite value $M$. Both $N$ and $M$ had to be increased until the examined results no longer changed with further increases, ensuring convergence is achieved. This procedure had to be performed for each parameter regime separately. In this Section, we demonstrate how that convergence analysis looks like in practice for the total number of phonons $M$. Convergence analysis with respect to the number of lattice sites $N$ will be presented in Sec.~\ref{sec:finite_size}. Before doing all of this, let us first address one theoretical question.

\medskip\noindent
Within the QT method, why is it justified to truncate the total number of phonons to be less or equal to $M$, given that the Hilbert space is inherently infinite-dimensional? In other words, why can the wave function $|\psi \rangle$, used for stochastic trace estimation, be represented as a finite-dimensional column vector in the coordinate representation, even though QT requires $|\psi\rangle$ to be constructed as a linear combination of an infinite number of basis vectors $|i\rangle$, with randomly chosen numbers $c_n$ as coefficients $|\psi\rangle = \sum_n c_n |n\rangle$; see Sec.~II of the main text. The justification lies in the fact that the relevant quantity in Eqs.~(10) and (11) of the main text is not $|\psi\rangle$ directly, but rather $|\psi_\beta\rangle = e^{-\beta H/2} |\psi\rangle$ and $|\phi_\beta\rangle = j e^{-\beta H/2} |\psi\rangle$. Therefore, we see that the the contribution corresponding to basis vectors with large number of phonons is suppressed by the Boltzmann factor $e^{-\beta H/2}$.

Let us now return to the main question of how the convergence with respect to $M$ is achieved in practice. A straightforward procedure would consist of repeating the QT calculation for $C_{jj}(t)$, using progressively larger values of $M$, until convergence is reached. However, this is actually not necessary. As an alternative, we can go back to Eq.~(11) from the main text, and analyze the scalar products in the numerator and denominator: in each of those scalar products, for each $t$, we can extract contributions corresponding to terms with total number of phonons less or equal to different values of $M=0,1,2\dots$. In such a way, repeating calculation for various $M$ is not needed, since a single QT calculation is sufficient. Usually, we apply the described technique  for somewhat smaller lattice $N$, estimate the necessary number of phonons per site $ M / N$, and then conclude that the total number of phonons for any larger lattice then just needs to be scaled appropriately $(M / N) N_\mathrm{larger}$.

\begin{figure}[!h]
  \includegraphics[width=6.5in,trim=0cm 0cm 0cm 0cm]{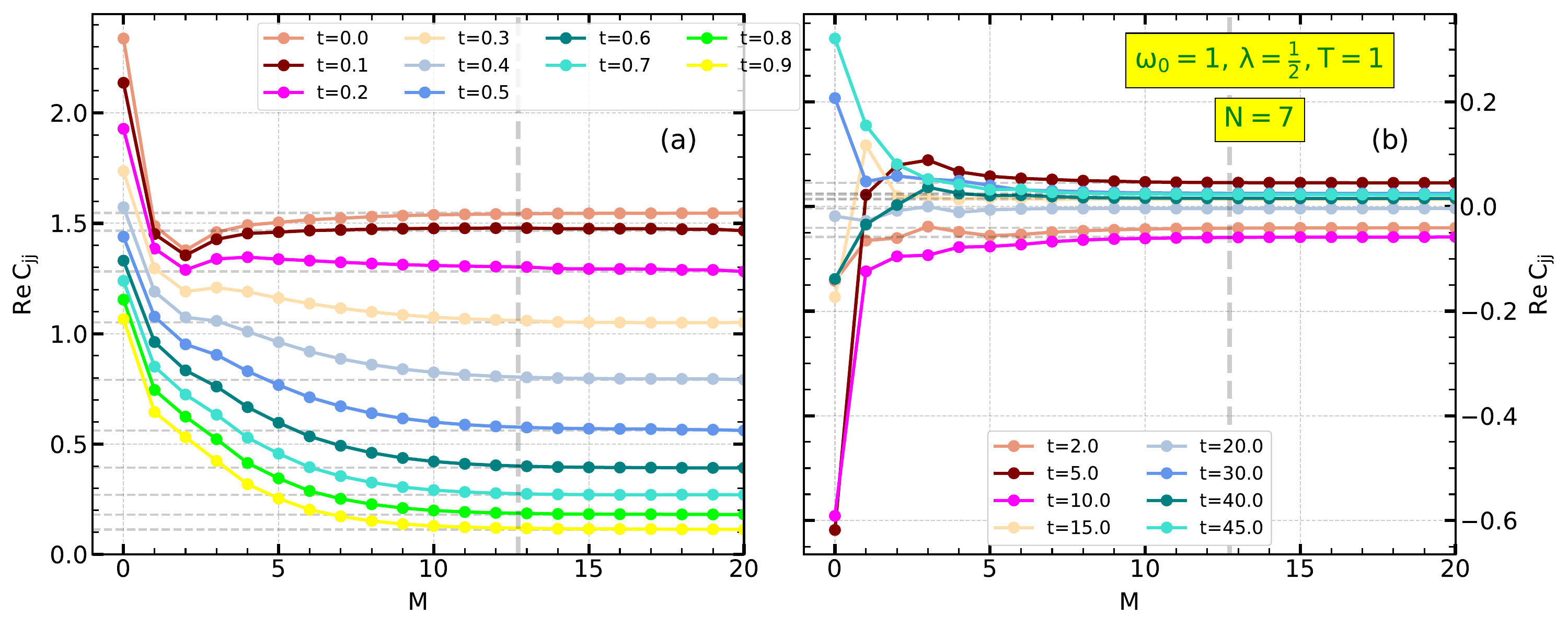} 
 \caption{Analyzing the convergence of the real part of the current current correlation function $\mathrm{Re}\,C_{jj}(t)$, with respect to the number of phonons $M$ taken in the Hilbert space for (a) small $t$ (b) large $t$. The tick vertical dashed line represents our estimation of the number of phonons one should take for the results to be considered converged for all times $t$. 
 The results are shown in the regime $\omega_0=1$, $\lambda=1/2$, $T=1$ on a lattice with $N=7$ sites.
 }
 \label{Fig:rezim_10}
\end{figure}

\begin{figure}[!h]
  \includegraphics[width=6.5in,trim=0cm 0cm 0cm 0cm]{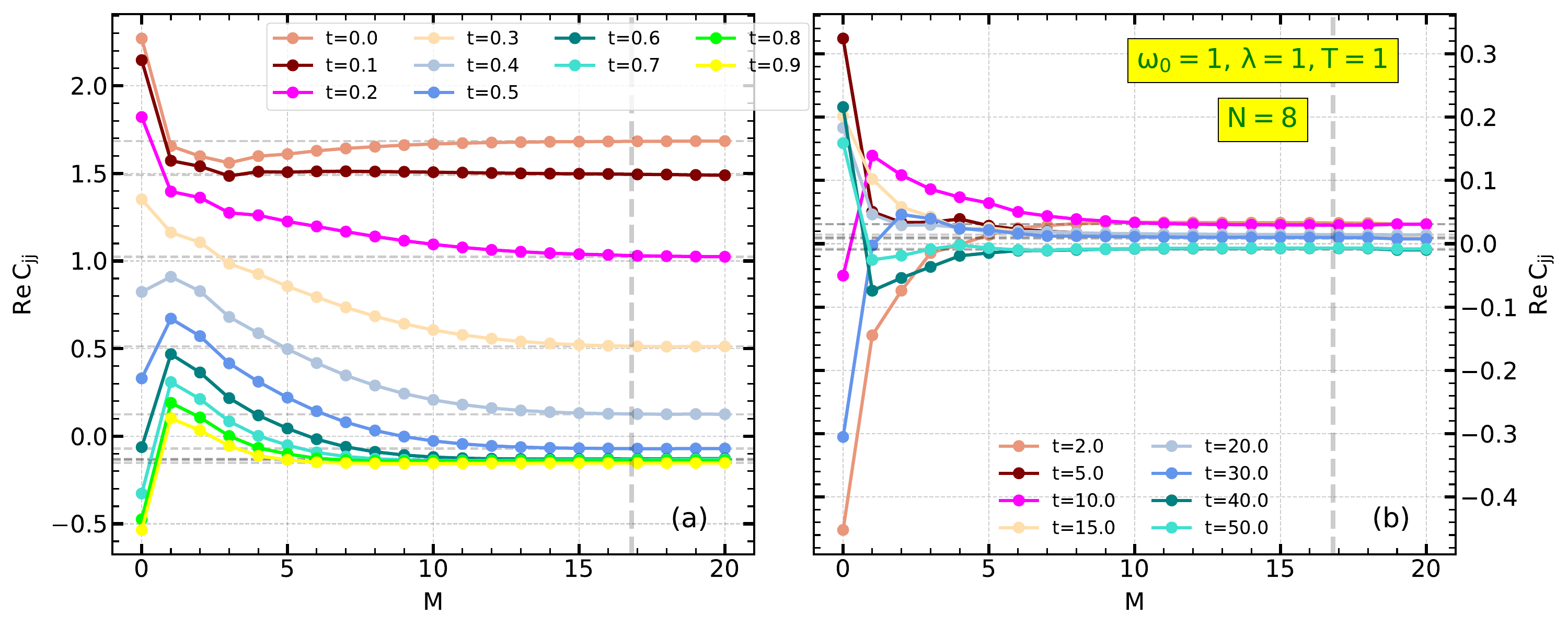} 
 \caption{Analyzing the convergence of the real part of the current current correlation function $\mathrm{Re}\,C_{jj}(t)$, with respect to the number of phonons $M$ taken in the Hilbert space for (a) small $t$ (b) large $t$. The tick vertical dashed line represents our estimation of the number of phonons one should take for the results to be considered converged for all times $t$. 
 The results are shown in the regime $\omega_0=1$, $\lambda=1$, $T=1$ on a lattice with $N=8$ sites.
 }
 \label{Fig:rezim_14}
\end{figure}

\begin{figure}[!h]
  \includegraphics[width=6.8in,trim=0cm 0cm 0cm 0cm]{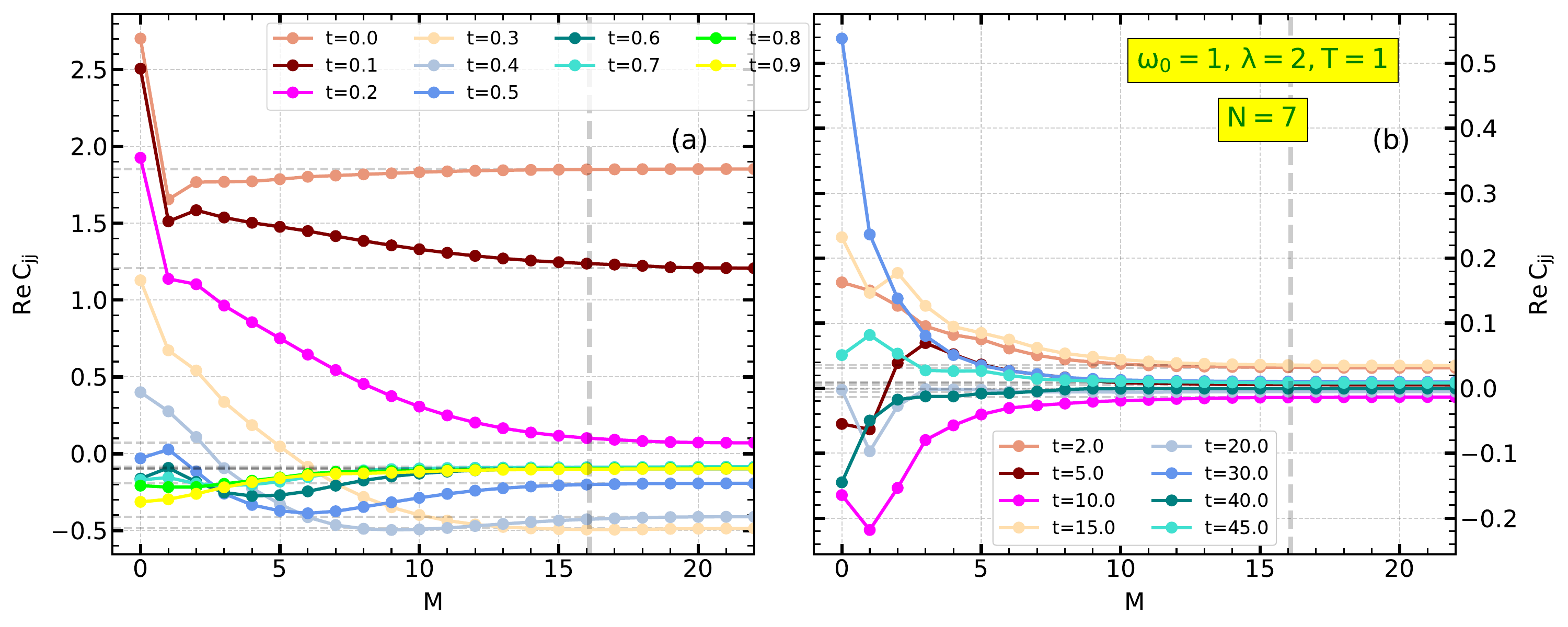} 
 \caption{Analyzing the convergence of the real part of the current current correlation function $\mathrm{Re}\,C_{jj}(t)$, with respect to the number of phonons $M$ taken in the Hilbert space for (a) small $t$ (b) large $t$. The tick vertical dashed line represents our estimation of the number of phonons one should take for the results to be considered converged for all times $t$. 
 The results are shown in the regime $\omega_0=1$, $\lambda=2$, $T=1$ on a lattice with $N=7$ sites.
 }
 \label{Fig:rezim_18}
\end{figure}

In Figs.~\ref{Fig:rezim_10},~\ref{Fig:rezim_14}~and~\ref{Fig:rezim_18}, we show how this analysis looks like for the regimes corresponding to top panels in Figs.~1,~2~and~3 of the main text. For example, in Fig.~\ref{Fig:rezim_10} we see that the convergence of $C_{jj}$ for all times $t$ is achieved for $M \approx 12$. This calculation was performed on a lattice with $N=7$ sites. Hence, for the lattice with $N=11$ sites, which we analyzed in the main text, one should use $M\approx 11\cdot (12/7) \approx 18.86$, which we rounded up to $M=20$. This is why $M=20$ was used in the main text.

It is interesting to note that the required number of phonons for convergence, as shown in Figs.~\ref{Fig:rezim_10},~\ref{Fig:rezim_14}, and~\ref{Fig:rezim_18}, appears to be smaller for larger times $t$. This trend is observed across all regimes analyzed in the main text. This is likely a consequence of the fact that the states with a large number of phonons correspond to higher energies, causing their phase to change rapidly as a function of $t$, particularly at large times, after the time evolution operator $e^{-i H t}$ is applied; see Eqs.~(12) and~(13) in the main text.

We note that while Figs.~\ref{Fig:rezim_10},~\ref{Fig:rezim_14}, and~\ref{Fig:rezim_18} illustrate only the dependence of $\mathrm{Re}\,C_{jj}(t)$ on $M$, a similar analysis was also conducted independently for both the numerator and denominator of Eq.~(11) from the main text. The reason why we do this, especially for the denominator, is because we want to make sure that the truncated Hilbert space can encompass all non negligible components of $| \psi_\beta \rangle$. This is important because the truncation of the Hilbert space can introduce error not only when calculating the scalar products (see Eq.~(11) of the main text), but also when performing time evolution of wave functions (see Eqs.~(12)~and~(13) of the main text).


%
\clearpage
\section{Relative error} \label{sec:rel_error}

\begin{figure}[t!]
  \centering
  \hypertarget{fig:deff}{} 
  \includegraphics[width=\textwidth]{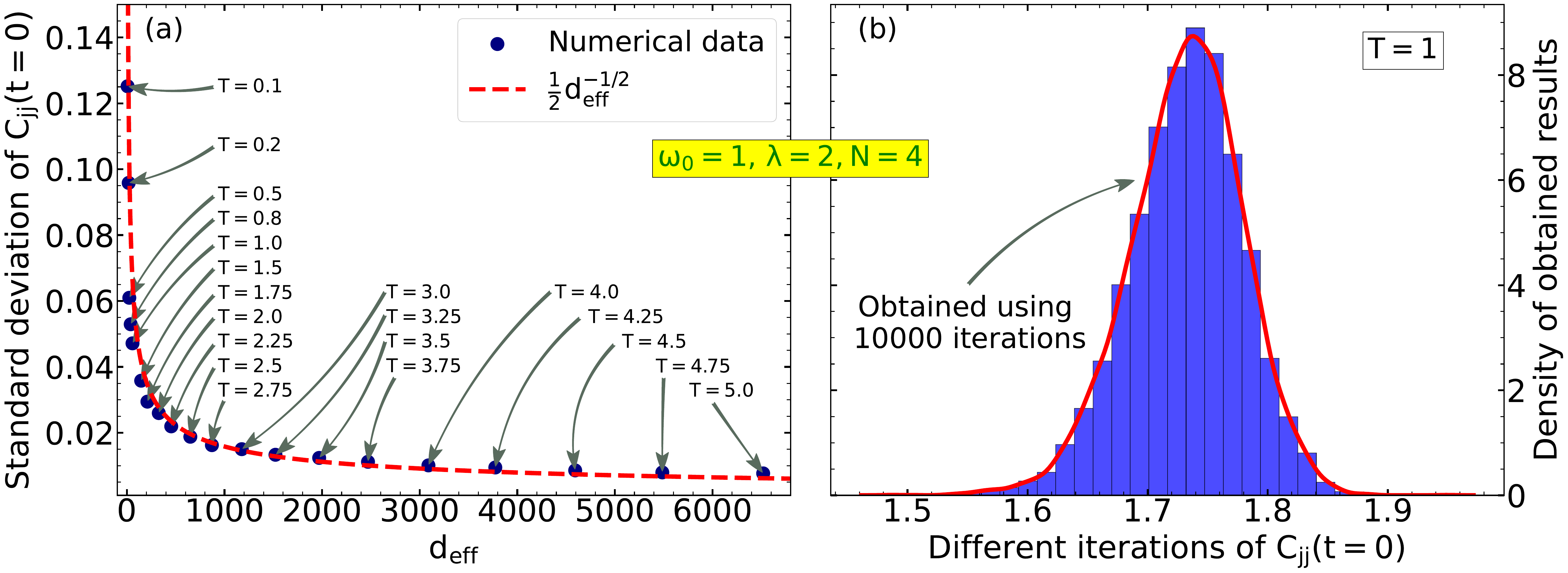}
  \caption{In the regime $\omega_0=1$, $\lambda=2$, $N=4$ we show (a) numerical results for the standard deviation of $C_{jj}(t=0)$ (i.e., estimated error for $C_{jj}(t=0)$ when it is calculated  using only single random vector in stochastic trace approximation) as a function of the effective dimensionality of the Hilbert space $d_{eff}$. In addition we also plot the function $\frac{1}{2} d_{eff}^{-1/2}$. (b) histogram that illustrates (for $T=1$) how frequently different results for $C_{jj}(t=0)$ appear when calculation, using only a single random vector, is repeated many times.}
  \label{fig:deff}
\end{figure}

In several references \cite{SM_2006_Weiss, SM_2013_Prelovsek} it is noted that the relative error one would obtain when using stochastic trace estimation for calculating some quantity, given by Eq.~(10) of the main text, scales as $\mathcal{O}(1/ \sqrt{R \, d_{eff}})$. Here, $R$ is the number of random vectors used in the stochastic trace approximation, while $d_{eff}$ is the effective dimension of the Hilbert space, given by  \( d_{eff} = \mathrm{Tr}[e^{-\beta(H-E_0)}] \), where $E_0$ is the ground state energy. It is clear that in the expression $\mathcal{O}(1/ \sqrt{R \, d_{eff}})$, the factor $1/\sqrt{R}$ follows directly from the central limit theorem. However, derivations of the factor $1/\sqrt{d_{eff}}$ that are commonly found in the literature \cite{SM_2006_Weiss, SM_2013_Prelovsek} are either not mathematically rigorous, or it is not completely clear that the assumptions used in the derivations are satisfied in the case of the current-current correlation function in electron-phonon systems. Here we will numerically check whether the error in the QT approach really scales as  $1/\sqrt{d_{eff}}$, in the case of the following quantity
\begin{equation}
    C_{jj}(t=0) = \frac{\mathrm{Tr}[e^{-\beta H} j^2]}{\mathrm{Tr}[e^{-\beta H} ]} = 
     \frac{\mathrm{Tr}[e^{-\beta H} j^2]}{Z}.
\end{equation}
For the sake of convenience in our numerical analysis, we first focus on the regime expected to exhibit a relatively large effective number of dimensions \( d_{eff} \). As an example, we consider the case of a relatively strong interaction \( \lambda = 2 \), a phonon frequency \( \omega_0 = 1 \), and a relatively high temperature \( T = 5 \). For this regime, we calculate \( C_{jj}(t=0) \) and the partition function \( Z \) for a lattice with \( N = 4 \) sites, using a sufficiently large number of phonons \( M \) to ensure full convergence (essentially treating \( M \to \infty \)). This calculation is then repeated many times to reliably compute \( Z \)\footnote{$Z$ is calculated using stochastic trace approximation, which is guaranteed to be accurate if we use sufficient number $R$ of random vectors, due to the error scaling $1/\sqrt{R}$ which we already noted is rigorous mathematical result.} and generate a statistical distribution (histogram) of the obtained values of \( C_{jj}(t=0) \). From this, we compute the standard deviation of \( C_{jj}(t=0) \), which serves as an estimate of the error in a single random iteration. In addition, we also calculate the effective dimensionality of the Hilbert space, as \( d_{eff} = \mathrm{Tr}[e^{-\beta(H-E_0)}] = e^{\beta E_0} \cdot Z \), where \( E_0 \) is the ground state energy. This entire procedure is then repeated for different temperatures, ranging from \( T = 5.0 \) to \( T = 0.1 \). The rationale behind this is that, at lower temperatures, fewer phononic excitations are relevant, leading to a smaller effective dimension of the Hilbert space \( d_{eff} \), enabling us to examine how the standard deviation of \( C_{jj}(t=0) \) scales as a function of \( d_{eff} \). The main results are presented in Fig.~\hyperref[fig:deff]{\ref{fig:deff}(a)}, while in Fig.~\hyperref[fig:deff]{\ref{fig:deff}(b)} we illustrate (on the example $T=1$, $\omega_0=1$, $\lambda=2$, $N=4$) the statistics of different possible values of $C_{jj}(t=0)$ obtained using stochastic trace approximation with only a single random vector. As shown in panel (a), the numerical data clearly demonstrate that the error in the calculated quantity is indeed proportional to \( 1/\sqrt{d_{eff}} \), as expected.
%
%

In table~\ref{tab:aaa} we show the dimension and effective dimension of the Hilbert space for every regime we examined in the main part of the text.
\begin{table}[ht] 
\centering
\begin{tabular}{c|c|c|c|c|c|c}
\hline
$\omega_0$ & $\lambda$ & $T$ & $N$ & $M$ & $d$ & $d_\mathrm{eff}$ \\
\hline
1.0 & 0.5 & 1.0 & 11 & 20 & 931395465 & 1485.56 \\
1.0 & 0.5 & 2.0 & 8 & 34 & 944241480 & 16312.7 \\
1.0 & 0.5 & 5.0 & 6 & 70 & 1311713640 & 260151.0 \\
\hline
1.0 & 1.0 & 1.0 & 10 & 21 & 443521650 & 1194.19 \\
1.0 & 1.0 & 2.0 & 7 & 33 & 130504920 & 6941.57 \\
1.0 & 1.0 & 5.0 & 6 & 70 & 1311713640 & 284888.0 \\
\hline
1.0 & 2.0 & 1.0 & 10 & 23 & 925610400 & 1812.83 \\
1.0 & 2.0 & 2.0 & 7 & 34 & 157373580 & 8960.15 \\
1.0 & 2.0 & 5.0 & 6 & 70 & 1311713640 & 319347.0 \\
1.0 & 2.0 & 10.0 & 5 & 125 & 1431218880 & 1233830.0 \\
\hline
0.333 & 0.5 & 1.0 & 7 & 45 & 936491920 & 48403.9 \\
0.333 & 0.5 & 5.0 & 5 & 130 & 1733501385 & 6636720.0 \\
\hline
0.333 & 1.0 & 1.0 & 7 & 48 & 1420494075 & 77565.8 \\
\hline
0.333 & 2.0 & 1.0 & 7 & 48 & 1420494075 & 92830.4 \\
\hline
\end{tabular}
\caption{The dimension $d$ and effective dimension $d_{eff}$ of the Hilbert space for every regime we examined in the main part of the text.}
\label{tab:aaa}
\end{table}

\section{Convergence with respect to the number of lattice sites $N$}%
\label{sec:finite_size}

One of the ways to analyze whether the calculated frequency-dependent mobility $\mu(\omega)$ has significant finite-size effects is to check whether the optical sum rule is satisfied; see Eqs.~(7)~and~(8) from the main part of the manuscript. Although this approach offers valuable insight, strictly speaking, satisfying the optical sum rule is only a necessary condition for the results to be representative of the thermodynamic limit. A more rigorous and direct approach is to simply calculate the results for different lattice sizes \( N \) and continue increasing \( N \) until \( \mu(\omega) \) fully converges. This is exactly what we will be doing here.
%

%
Before presenting the results, we note that for sufficiently small lattice sizes \( N \), a single random vector in the stochastic trace approximation is insufficient for obtaining accurate results;  see Sec.~\ref{sec:rel_error} and Sec.~IIC of the main text. Therefore, while we typically use  \( R = 1 \) random vector for the largest lattice sizes, a larger number of random vectors $R$ is employed for smaller lattices.

\subsection{$\omega_0=1$, $\lambda=\frac{1}{2}$} \label{subsec:weakcop}

\begin{figure}[!h]
  \includegraphics[width=6.8in,trim=0cm 0cm 0cm 0cm]{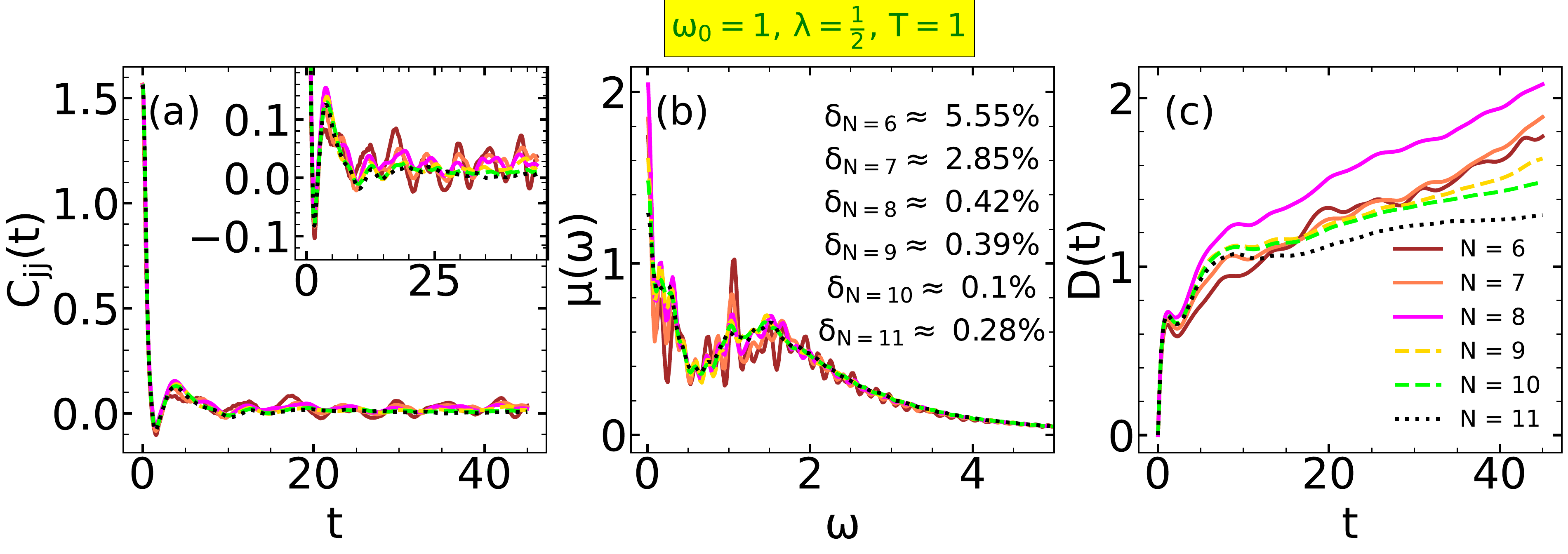} 
 \caption{Results for (a) the real part of the current-current correlation function $\mathrm{Re}\,C_{jj}(t)$ (b) frequency dependent mobility $\mu(\omega)$ (c) time dependent diffusion constant $D(t)$, in the regime $\omega_0=1, \lambda=\frac{1}{2}, T=1$ for different number of lattice sites $N$. The number of random vectors, used for the calculation of the results on lattices of various sizes, are $R_{N=11}=1$, $R_{N=10}=10$, $R_{N=9}=10$, $R_{N=8}=50$, $R_{N=7}=120$, $R_{N=6}=330$.
 }
 \label{Fig:a1}
\end{figure}

By examining the results in Fig.~\ref{Fig:a1}, and by inspecting how various cutoffs in time (for $C_{jj}(t)$) influence the behavior of $\mu(\omega)$ (not shown), we see that \(\mu(\omega)\), for \(\omega \gtrsim 0.1\), converges already for \(N = 9\). However, achieving full convergence at \(\mu(\omega = 0)\) is significantly more challenging. This is reflected in the discrepancies between the diffusion constants \(D(t)\) at large times for different \(N\), as a consequence of the Einstein relation
\begin{equation}
    \mu(\omega = 0) = \frac{D(t\to\infty)}{T}.
\end{equation}
Numerical results for the diffusion constants \( D(t) \) are shown in Panel~(c). It is clear that for \( N \leq 10 \), a significantly longer time propagation is needed to reach \( D(t \to \infty) \), i.e., to reach saturation of the diffusion constant. However, for larger lattices, \( D(t) \) appears to saturate more quickly. Nevertheless, even for $N=11$, $D(t)$ is not fully saturated at Panel~(c). Therefore, we expect that the correct result for $D(t\to\infty)$ (and thus $\mu(\omega=0)$) is somewhat different than shown in Fig.~\ref{Fig:a1} (and thus closer to the HEOM result shown in Fig.~1($(\mathrm{a}_2)$ of the main text). We note that the HEOM result for $D(t)$, used in the main text, was propagated up to much longer time $t$ (and also for somewhat larger lattice $N=13$), so the saturation of $D(t)$ is much better. It is thus expected that HEOM is, in this case, more accurate than QT.
\par\medskip\noindent 
{\textbf{Side Note: }} As we see from Fig.~\hyperref[fig:deff]{\ref{Fig:a1}(b)}, as we increase the lattice size, the accuracy with which the optical sum rule is satisfied (measured by how close the parameter $\delta$ is to zero; see Eq.~(8) from the main text) increases. There is only one exception: $\delta_{N=11} \approx 0.28 \%$ which is slightly larger than $\delta_{N=10} \approx 0.1 \%$. Regarding this, we would like to point out two things: i) both of these cases ($N=10$ and $N=11$) satisfy the optical sum rule fairly well. ii) for this regime, the effective dimension of the Hilbert space $d_{eff}$ is not very large (see Table~\ref{tab:aaa}), so there probably are some slight incurabilities that arose because we used only a single random vector $R=1$ for $N=11$. On the other hand, for $N=10$ we used $R=10$ random vectors.
\par
\begin{figure}[!h]
  \includegraphics[width=6.8in,trim=0cm 0cm 0cm 0cm]{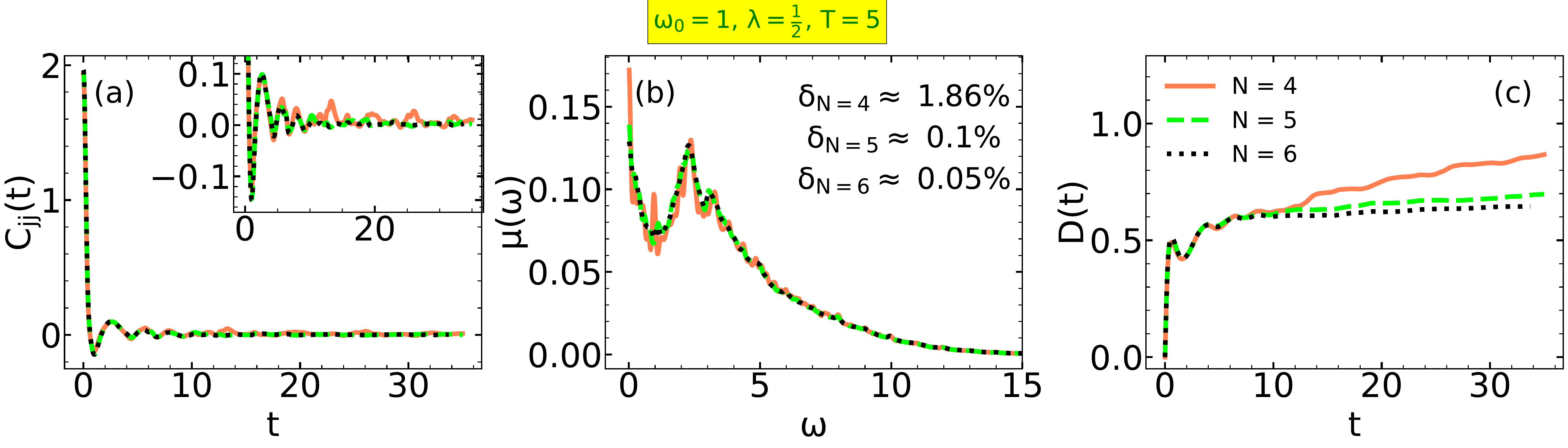} 
 \caption{Results for (a) the real part of the current-current correlation function $\mathrm{Re}\,C_{jj}(t)$ (b) frequency dependent mobility $\mu(\omega)$ (c) time dependent diffusion constant $D(t)$, in the regime $\omega_0=1, \lambda=\frac{1}{2}, T=5$ for different number of lattice sites $N$. The number of random vectors, used for the calculation of the results on lattices of various sizes, are $R_{N=6}=1$, $R_{N=5}=2$, $R_{N=4}=65$.
 }
 \label{Fig:a2}
\end{figure}

\medskip
Analogous conclusions can also be drawn for Fig.~\ref{Fig:a2}. In this case, however, a much better saturation (in the case of QT method) of the diffusion constant $D(t)$ is observed. In addition, we see that $N=5$ and $N=6$ results practically coincide, so the finite-size effects are minimal. Some difference  between HEOM and QT predictions for $D(t\to\infty)$ is however observed, as shown in the inset of Fig.~1($\mathrm{c}_2$) of the main text. This is a consequence of the inaccuracy of the HEOM solution (see the inset of Fig.~1($\mathrm{c}_1$) and the associated text) at intermediate timescales $t \gtrsim 2$. Hence, we conclude that QT is more precise in this case. 

\subsection{$\omega_0=1$, $\lambda=1$} \label{sec:interM_co}

\begin{figure}[!h]
  \includegraphics[width=6.8in,trim=0cm 0cm 0cm 0cm]{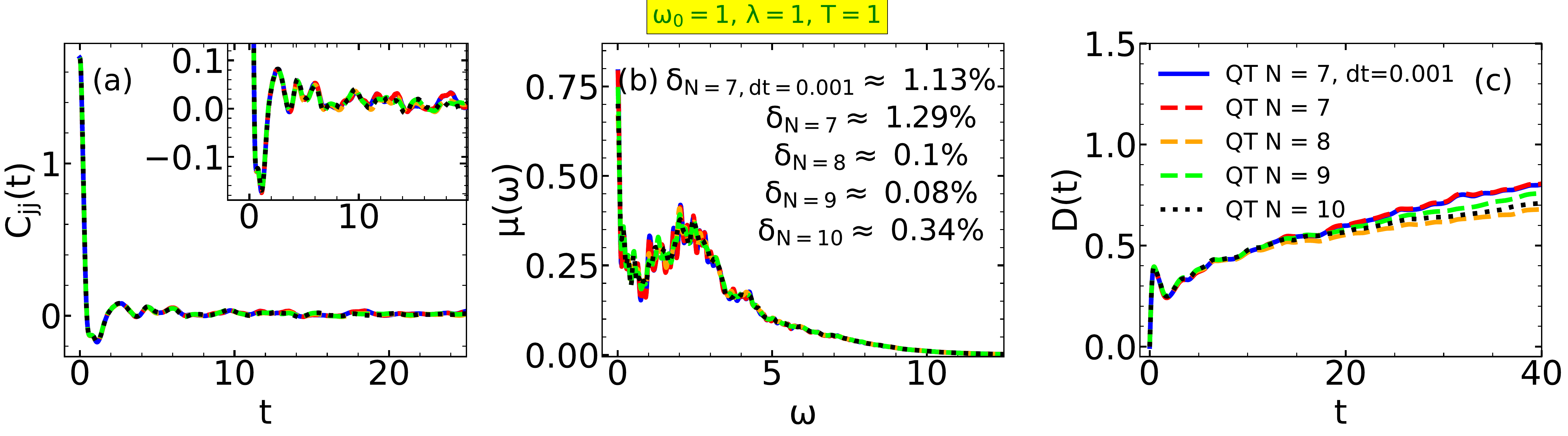} 
 \caption{Results for (a) the real part of the current-current correlation function $\mathrm{Re}\,C_{jj}(t)$ (b) frequency dependent mobility $\mu(\omega)$ (c) time dependent diffusion constant $D(t)$, in the regime $\omega_0=1, \lambda=1, T=1$ for different number of lattice sites $N$. For $N=7$ we also show the results for two different time steps ($dt=0.01$ and $dt=0.001$; the results for all other lattice sizes were obtained using  $dt=0.01$). The number of random vectors, used for the calculation of the results on lattices of various sizes, are $R_{N=10}=1$, $R_{N=9}=15$, $R_{N=8}=30$, $R_{N=7}=100$, $R_{N=7, dt=0.001}=100$.
 }
 \label{Fig:a7}
\end{figure}

For $\omega_0=1, \lambda=1$ the conclusions are analogous to the ones we have already drawn for $\omega_0=1, \lambda=\frac{1}{2}$ in Sec.~\ref{subsec:weakcop}: As seen from Fig.~\ref{Fig:a7}, although for $\omega \gtrsim 0.15$ the thermodynamic limit is nearly reached for $N \geq 8$, the diffusion constant $D(t)$ does not sufficiently saturate within the shown time scale, preventing a reliable determination of the DC mobility. Here, we have also explicitly demonstrated that the absence of saturation of $D(t)$ is not a consequence of the Runge-Kutta timestep: for $N=7$ we have checked this using two different timesteps,  $dt=0.01$ and $dt=0.001$. In addition, as always, we checked that the results have fully converged with respect to the total number of phononic excitations (see Fig.~\ref{Fig:rezim_14}). Having in mind the conclusions drawn in Sec.~\ref{subsec:weakcop}, we can thus expect that the absence of saturation of $D(t)$ is probably due to the finite-size effects.

As before, for higher temperature, $T=5$, a much better saturation of $D(t)$ can be observed. In addition, results for $N=6$ and $N=5$ are quantitatively quite similar, so we can expect that the calculated DC mobility (and the whole $\mu(\omega)$) should be accurate.
\begin{figure}[!h]
  \includegraphics[width=6.8in,trim=0cm 0cm 0cm 0cm]{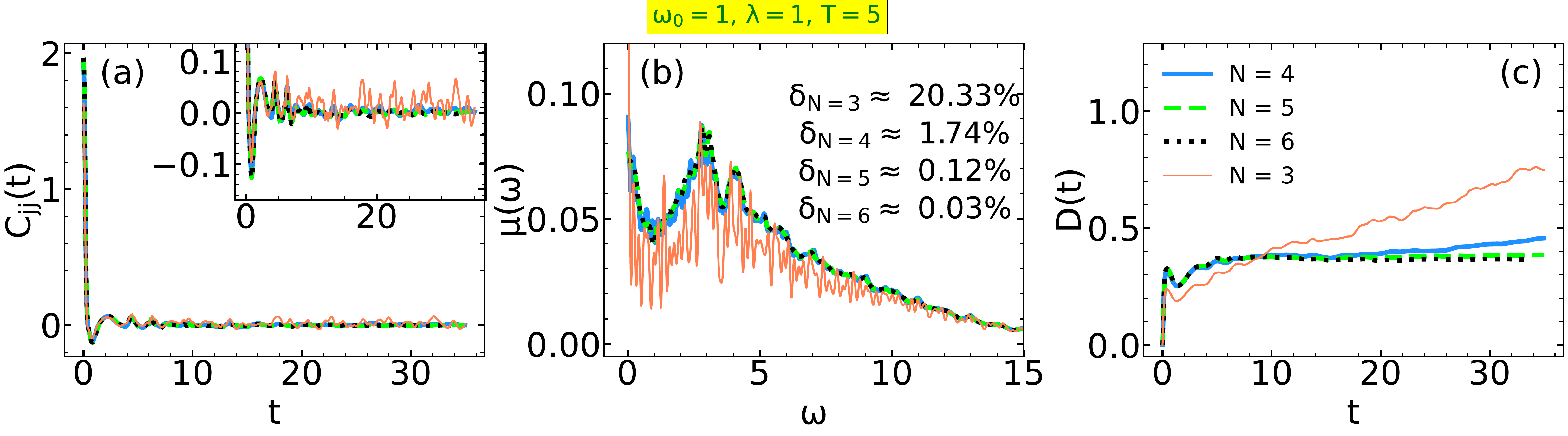} 
 \caption{Results for (a) the real part of the current-current correlation function $\mathrm{Re}\,C_{jj}(t)$ (b) frequency dependent mobility $\mu(\omega)$ (c) time dependent diffusion constant $D(t)$, in the regime $\omega_0=1, \lambda=1, T=5$ for different number of lattice sites $N$. The number of random vectors, used for the calculation of the results on lattices of various sizes, are $R_{N=6}=1$, $R_{N=5}=2$, $R_{N=4}=70$, $R_{N=3}=400$.
 }
 \label{Fig:a3}
\end{figure}

\subsection{$\omega_0=\frac{1}{3}, \lambda=\frac{1}{2}$}

%
\begin{figure}[!h]
  \includegraphics[width=6.8in,trim=0cm 0cm 0cm 0cm]{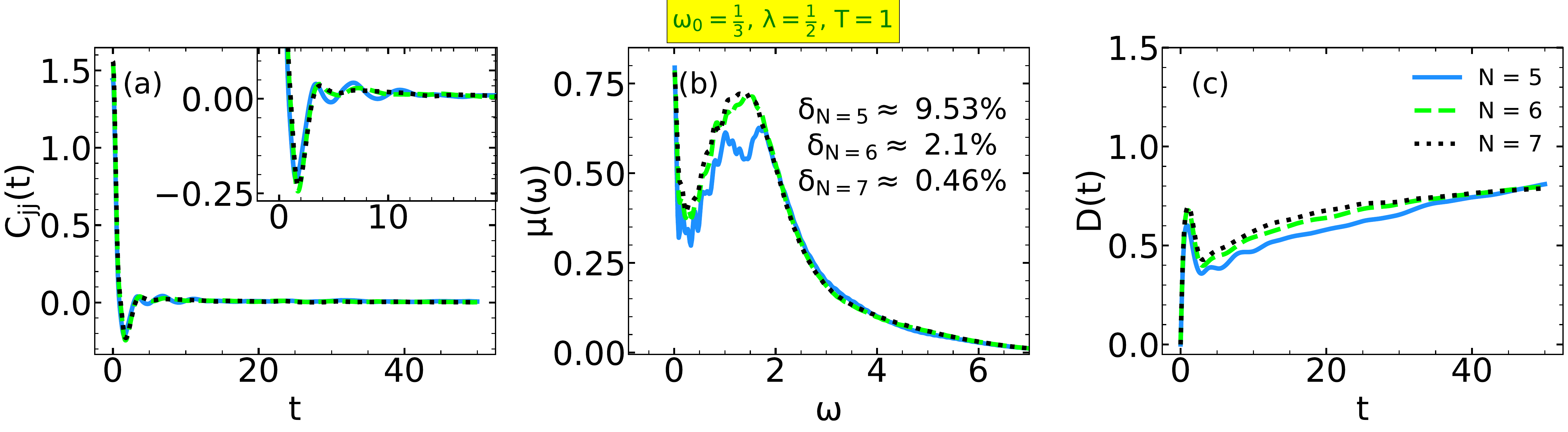} 
 \caption{Results for (a) the real part of the current-current correlation function $\mathrm{Re}\,C_{jj}(t)$ (b) frequency dependent mobility $\mu(\omega)$ (c) time dependent diffusion constant $D(t)$, in the regime $\omega_0=\frac{1}{3}, \lambda=\frac{1}{2}, T=1$ for different number of lattice sites $N$. The number of random vectors, used for the calculation of the results on lattices of various sizes, are $R_{N=7}=1$, $R_{N=6}=7$, $R_{N=5}=200$.
 }
 \label{Fig:a4}
\end{figure}
\begin{figure}[!h]
  \includegraphics[width=6.8in,trim=0cm 0cm 0cm 0cm]{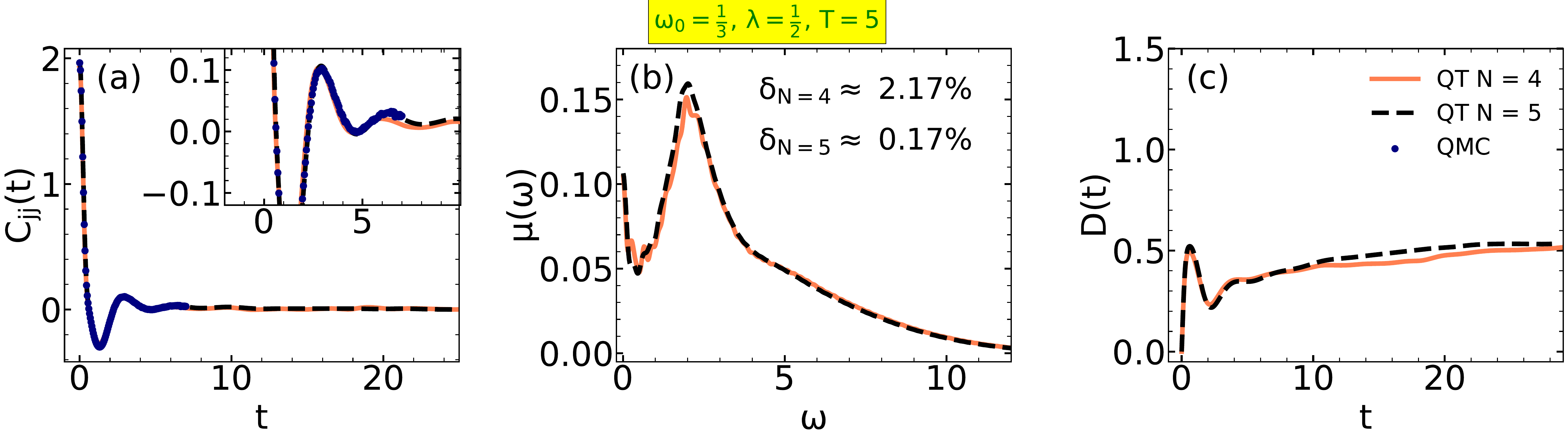} 
 \caption{Results for (a) the real part of the current-current correlation function $\mathrm{Re}\,C_{jj}(t)$ (b) frequency dependent mobility $\mu(\omega)$ (c) time dependent diffusion constant $D(t)$, in the regime $\omega_0=\frac{1}{3}, \lambda=\frac{1}{2}, T=5$ for different number of lattice sites $N$. The number of random vectors, used for the calculation of the results on lattices of various sizes, are $R_{N=5}=1$, $R_{N=4}=5$.
 }
 \label{Fig:a5}
\end{figure}

To avoid redundancy, we will not discuss the results in this regime, as everything is analogous to the analysis we conducted in Secs.~\ref{subsec:weakcop}~and~\ref{sec:interM_co}.

\section{Additional results in the strong coupling regime $\lambda=2$}
\poc
In the main part of the text, in Fig.~3, we presented and analyzed the current-current correlation function $C_{jj}(t)$, the optical conductivity $\mu(\omega)$, and the time-dependent diffusion constant $D(t)$ in the strong coupling regime $\lambda=2$ for $\omega_0=1$ and $T=1,10$. These results were shown within QT, QMC and DMFT methods, while HEOM results were missing due to the problems with convergence with respect to the so-called maximum hierarchy depth. Here, we supplement those results by also adding the results for $T=2$ and $T=5$. As these results were already discussed in the main text, we only show the results without additional comments.


%
\begin{figure}[!h]
  \includegraphics[width=7.0in,trim=0cm 0cm 0cm 0cm]{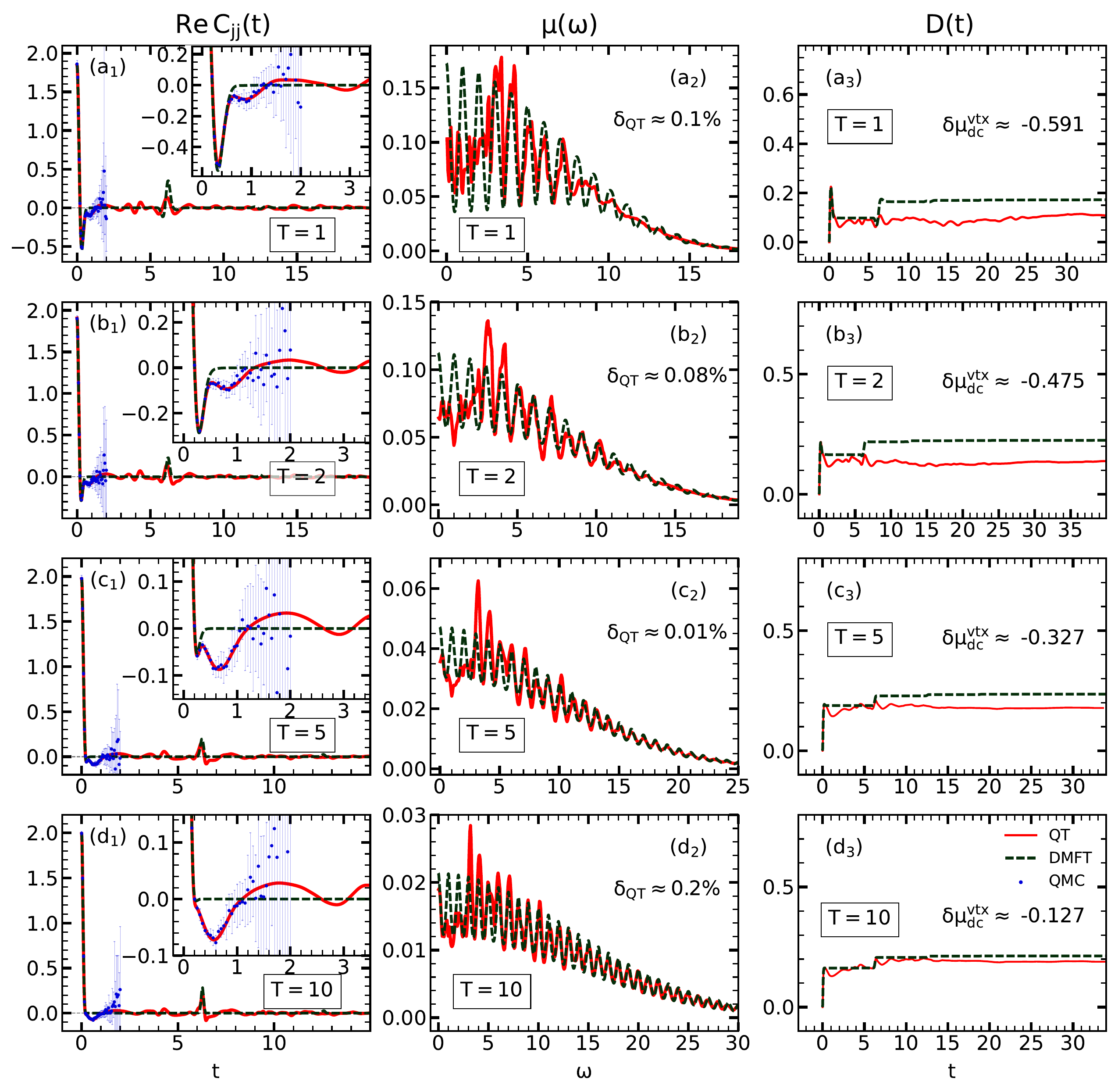} 
 \caption{Comparison of $(\mathrm{a}_1)$--$(\mathrm{d}_1)$ the real part of the current-current correlation function $C_{jj}(t)$ $(\mathrm{a}_2)$--$(\mathrm{d}_2)$ frequency dependent mobility $\mu(\omega)$ $(\mathrm{a}_3)$--$(\mathrm{d}_3)$ time dependent diffusion constant $D(t)$. The results are calculated using QT, QMC and DMFT, in the strong coupling regime $\omega_0 = 1$, $\lambda = 2$, for $T=1,2, 5, 10$. In Panels $(\mathrm{a}_2)$--$(\mathrm{d}_2)$ we show the relative accuracy $\delta$ with which the optical sum rule is satisfied within QT, in $(\mathrm{a}_3)$--$(\mathrm{d}_3)$ we show the significance of vertex correction for DC mobility $\delta \mu_{dc}^\mathrm{vtx}$, given by Eq.~(18) from the main text, while the insets of $(\mathrm{a}_1)$--$(\mathrm{d}_1)$ show the zoomed-in portions of the corresponding panels.  The total number of phonons and lattice sizes used are: $(\mathrm{a}_1)$--$(\mathrm{a}_3)$ $(M^\mathrm{QT}, N^\mathrm{QT}) = (23, 10)$; $(\mathrm{b}_1)$--$(\mathrm{b}_3)$ $(M^\mathrm{QT}, N^\mathrm{QT}) = (34, 7)$; $(\mathrm{c}_1)$--$(\mathrm{c}_3)$ $(M^\mathrm{QT}, N^\mathrm{QT}) = (70, 6)$; $(\mathrm{d}_1)$--$(\mathrm{d}_3)$ $(M^\mathrm{QT}, N^\mathrm{QT}) = (125, 5)$. QMC results were always obtained for $N^\mathrm{QMC}=10$, and they are shown with their corresponding statistical error bars. DMFT results were calculated directly in the thermodynamic limit $N^\mathrm{DMFT} \to \infty$.
 }
 \label{Fig:strong_coupling}
\end{figure}

\clearpage
\newpage

%

\end{document}